\documentclass[english,aps,preprint,nofootinbib]{revtex4}
\usepackage[T1]{fontenc}
\usepackage[latin9]{inputenc}
\usepackage{color}
\usepackage{amsbsy}
\usepackage{amssymb}
\usepackage{esint}

\makeatletter
\definecolor{dyellow}{rgb}{1.,0.8,.0}
\definecolor{myblue}{rgb}{.1,.1,.7}
\definecolor{dcyan}{rgb}{.0,.6,.6}
\definecolor{dmagenta}{rgb}{0.6,0.0,0.6}
\definecolor{brown}{rgb}{0.6,0.2,0.}
\definecolor{darkblue}{rgb}{.0,.0,0.5}
\definecolor{darkred}{rgb}{0.75,0.0,0.0}
\definecolor{orange}{rgb}{1.,.6,.0}
\definecolor{dorange}{rgb}{0.8,.4,.0}
\definecolor{darkgreen}{rgb}{0.0,0.6,0.0}
\definecolor{purple}{rgb}{.4,.0,.4}
\definecolor{lightgray}{rgb}{.8,.8,.8}

\usepackage{babel}

\makeatother

\usepackage{babel}

\newcommand{\heq}{{\ {\hat=}\ }}
\newcommand{\del}{{\nabla}}

\begin{document}

\preprint{This line only printed with preprint option}

\title{Holographic Entropy Production}

\author{Yu Tian}

\email{ytian@ucas.ac.cn}

\affiliation{School of Physicas, University of Chinese Academy of Sciences, Beijing
100049, China\\
State Key Laboratory of Theoretical Physics, Institute of Theoretical
Physics, Chinese Academy of Sciences, Beijing 100190, China}

\author{Xiao-Ning Wu}

\email{wuxn@amss.ac.cn}

\affiliation{Institute of Mathematics, Academy of Mathematics and System Science,
CAS, Beijing 100190, China\\
State Key Laboratory of Theoretical Physics, Institute of Theoretical
Physics, Chinese Academy of Sciences, Beijing 100190, China}

\author{Hongbao Zhang}

\email{hzhang@vub.ac.be}

\affiliation{Theoretische Natuurkunde, Vrije Universiteit Brussel and The International
Solvay Institutes, Pleinlaan 2, B-1050 Brussels, Belgium}
\begin{abstract}
{The suspicion that gravity is holographic has been supported mainly by a variety of specific examples from
string theory. In this paper, we propose that such a holography can actually be observed in the context of Einstein's
gravity and at least a class of generalized gravitational theories, based on a definite holographic principle
where neither is the bulk space-time required to be asymptotically AdS nor the boundary to be located
at conformal infinity, echoing Wilson's formulation of quantum field theory.
After showing the general equilibrium thermodynamics from the corresponding holographic dictionary,
in particular, we provide a rather general proof of the equality between the entropy production on the boundary and the
increase of black hole entropy in the bulk, which can be regarded as strong support to this holographic principle.
The entropy production in the familiar holographic superconductors/superfluids is investigated
as an important example, where the role played by the holographic renormalization is explained.}
\end{abstract}
\maketitle
\tableofcontents{}

\section{Introduction}

{Evidence has accumulated} since the end of last century that
quantum gravity is holographic \cite{Hooft,Susskind}, i.e. quantum
gravity in a $(d+1)$-dimensional space-time region can be described
by some sort of quantum field theory on {the $d$-dimensional boundary of this region},
especially since the discovery of AdS/CFT correspondence \cite{Maldacena,Witten,GKP}
in the framework of (super)string theory. On one hand, nowadays there
have been many generalizations and/or applications of AdS/CFT correspondence,
such as most of the phenomenological models in AdS/CMT (condensed
matter theory), AdS/QCD and so on, which cannot be embedded in string
theory. On the other hand, besides the black hole thermodynamics \cite{Wald}
that inspires the proposition of holography, there are already various
hints from within the context of Einstein's gravity towards the speculation
that gravity is essentially holographic, where neither string theory
nor supersymmetry are involved. Here we would like list three of them
as follows.
\begin{itemize}
\item Brown-Henneaux's asymptotic symmetry analysis for three dimensional
gravity \cite{B H}.
\item Brown-York's surface tensor formulation of quasilocal energy and conserved
charges \cite{B Y}.
\item Bousso's covariant entropy bound \cite{Bousso}.
\end{itemize}
In particular, Brown-York's surface tensor formulation bears a strong
resemblance to the recipe in the dictionary for AdS/CFT correspondence,
and has actually been incorporated into the latter (or its generalizations).
{Holography could} have been explicitly implemented just in Einstein's
gravity, in fact, if one was brave enough to declare that Brown-York's
surface tensor is not only for the purpose of the bulk side but also
for some sort of system living on the boundary.

In AdS/CFT, the radial direction of the (asymptotic AdS) bulk space-time
corresponds to the energy scale of the dual field theory \cite{BKLT,Akhmedov,dBVV,S W}
and the change of radial coordinate $r$ is regarded as equivalent
to the corresponding renormalization group (RG) flow \cite{KSS,S S,B L,H P,FLR,Sin},
where the conformal boundary $r\to\infty$ is its ultra-violet (UV)
fixed point. Interestingly, from this point of view, the RG flows
of many important transport coefficients of the boundary theory (at
finite temperature) are trivial, which enables one to compute these
coefficients by the so-called black-hole ``membrane paradigm'' {\cite{I L}}.
Especially, it is proved that the ratio $\frac{\eta}{s}=\frac{1}{4\pi}$
of the shear viscosity $\eta$ to the entropy density $s$ does not
run with the RG flow, so the universality of this ratio in both the
black-hole membrane paradigm and the standard AdS/CFT follows.

In the above framework of the so-called holographic RG flow, physical
quantities can be defined on any constant $r$ surface (called the
finite cutoff surface), while their RG flows are obtained by changing
$r$. However, the finite cutoff surface itself is just a tool to
relate the conformal boundary $r\to\infty$ (UV) and the bulk black-hole
horizon $r\to r_{h}$ (IR), and no dual dynamical theory is directly
defined on this surface, until later Strominger et al \cite{Strominger,Strominger2}
establish hydrodynamics on the finite cutoff surface and {then}
discuss the fluid/gravity correspondence from this point of view.
The dual theory defined on the finite cutoff surface $r=r_{c}$ can
be regarded as the effective field theory at the energy scale corresponding
to $r_{c}$. In fact, the bulk space-time in this generalized framework
of holography can be either asymptotic AdS \cite{Cai,NTWL,Wang} or
not \cite{Strominger2,Skenderis2}, reflecting the fact that the dual
theory does not need to have a UV completion.

Both in the standard AdS/CFT at the conformal boundary and in the
generalized holography by Strominger et al, a holographic
interpretation of the entropy production of the boundary system in
non-equilibrium processes is an interesting problem. It has been
well established in AdS/CFT that a static black hole in the bulk is
dual to the boundary field theory at a thermal equilibrium state.
Then what happens when the bulk black hole is perturbed? From the
bulk point of view, the bulk perturbation will be eventually
absorbed by the black hole, leading to an increase of the area of
black hole, i.e., an increase of the black hole entropy \cite{I W}.
On the other hand, such a bulk perturbation will induce the
corresponding perturbation on the boundary, driving the boundary
system away from the original equilibrium state. But the dissipation
will bring the boundary system to a new equilibrium state with the
production of entropy. So a natural question is whether the entropy
production by such a dissipative (transport) process on the boundary
is equal to the increase of the black hole entropy in the bulk.
Actually this problem has been raised by Strominger et al in the
generalized holography \cite{Strominger}, but it still remains open until now.

So the main motivation of our paper is two-fold. On one hand, the
UV fixed point of the dual field theory, which has a conformal dynamics,
is not expected to be reached by experiments. Therefore, the generalized
holography, which we call the general bulk/boundary correspondence,
at a finite cutoff surface $r=r_{c}$ (corresponding to a finite energy
scale) is important, where the dual (effective) theory is non-conformal
in general. In order to study the general bulk/boundary correspondence
systematically, we {propose} a general holographic principle, which leads
to definite holographic dictionary on any cutoff surface. This dictionary
should include the known cases \cite{Strominger,Strominger2} as special
examples, and should be consistent with the standard AdS/CFT when
$r\to\infty$, if certain subtleties like the holographic renormalization%
\footnote{For a nice review of the holographic renormalization, see, e.g., Ref.\cite{Skenderis}.%
} are taken into account.

On the other hand, {as a support for our general bulk/boundary correspondence,}
we prove that the entropy production by the
transport processes on the boundary is exactly equal to the increase of the
black-hole entropy in the bulk an explicit construction of certain
conserved currents, which is rather involved in the case of coupled
transport processes. Since the holographic
picture of general non-equilibrium processes has difficulties from
both conceptual and technical aspects, we consider the near-equilibrium
cases here, which corresponds to linear perturbations of the background
bulk configuration. Then the discussion can be extended
to the usual holographic models such as holographic superconductors/superfluids
on the conformal boundary, after considering the holographic renormalization.
Even without knowing holography, such an equality, together with the
traditional black-hole membrane paradigm \cite{membrane}, can be
viewed as generalization of the well-established black-hole thermodynamics
to the black-hole ``hydrodynamics'' (see also Ref.\cite{Oz}).

The rest of our paper is structured as follows. In Section II, we
briefly review the basic idea of the general holographic principle,
the corresponding dictionary and its implementation in the static
case. In Section III, we present our proof of the above equality by
connecting the bulk with the boundary through the conserved current.
We then analyze the entropy production in holographic superconductors/superfluids
in Section IV. The last section is dedicated to some discussions on
our result.

\section{Holographic dictionary and its implementation in the equilibrium
thermodynamics}

Our starting point is the following (Euclidean) holographic principle
\begin{equation}
Z_{{\rm bulk}}[\bar{\phi}]=\int D\psi\exp(-I_{{\rm FT}}[\bar{\phi},\psi])\label{eq:Euclidean}
\end{equation}
for some quantum gravity theory with partition function $Z_{{\rm bulk}}[\bar{\phi}]$
on some bulk space-time region and the corresponding quantum field
theory with action $I_{{\rm FT}}[\bar{\phi},\psi]$ on its boundary,
which is the refined and generalized version of the original AdS/CFT
principle \cite{Witten,GKP}. Here the partition function $Z_{{\rm bulk}}[\bar{\phi}]$
is evaluated by fixing the boundary value of the bulk field $\phi$
to be $\bar{\phi}$, which acts as some background field on the boundary,
and $\psi$ denotes all the dynamical fields in the boundary theory,
which is integrated out to produce the partition function in the right
hand side of (\ref{eq:Euclidean}). To be more precise, if $\phi$
is the metric or form fields, then the pull back of $\phi$ to the
boundary is fixed to be $\bar{\phi}$. Infinitesimal variation of
$\bar{\phi}$ in (\ref{eq:Euclidean}) gives
\[
Z_{{\rm bulk}}[\bar{\phi}+\delta\bar{\phi}]=Z_{{\rm bulk}}[\bar{\phi}]\left\langle \exp\int_{{\rm bdry}}\delta\bar{\phi}O_{\phi}\sqrt{\bar{g}}d^{d}x\right\rangle _{\mathrm{FT}}
\]
with $\sqrt{\bar{g}}d^{d}x$ the standard volume element on the boundary
and
\[
O_{\phi}(x)=-\frac{1}{\sqrt{\bar{g}}}\frac{\delta I_{{\rm FT}}[\bar{\phi},\psi]}{\delta\bar{\phi}(x)}
\]
the ``dual field'', which should be understood as the corresponding
quantum operator in the expression of expectation value.

In the classical limit (or sometimes called the saddle point approximation),
the bulk partition function is given by
\[
Z_{{\rm bulk}}[\bar{\phi}]=\exp(-I_{\mathrm{bulk}}[\bar{\phi}])
\]
with $I_{\mathrm{bulk}}[\bar{\phi}]$ the on-shell action (Hamilton's
principal functional). So the above holographic principle leads to
\begin{equation}
-\frac{1}{\sqrt{g}}\frac{\delta I_{\mathrm{bulk}}[\bar{\phi}]}{\delta\bar{\phi}(x)}=\left\langle O_{\phi}(x)\right\rangle _{{\rm FT}},\label{eq:dictionary}
\end{equation}
where the left hand side is just the canonical momentum conjugate
to $\phi$ by virtue of the Hamilton-Jacobi equation regarding the
boundary as the ``time'' slice. Now turn to the Minkowskian signature.
The discussion in this case is similar to the above, but subtleties
arise when one further considers correlation functions \cite{Minkowski},
which does not concern us in the present paper. For the bulk being
(asymptotic) AdS space-time and the boundary tending to its conformal boundary,
it is well known that the dual field theory is a (local) CFT.\footnote{In this case,
some related discussions can be seen in Sec.\ref{sec:conformal}.}
But in more general cases, e.g. asymptotically flat bulk and/or boundary
at finite distance \cite{Strominger,Strominger2}, the dual theory
should be some effective field theory that is both non-local and non-conformal
\cite{non-local}, inspired by the well-known AdS/CFT interpretation
that the radial direction is related to renormalization group flow
of the dual theory. Although the details of the general dual theory
is unclear so far, macroscopic aspects of the general bulk/boundary
correspondence turn out to be universal and can be clearly understood,
which is part of the main motivation of this paper. In the macroscopic
point of view, the boundary theory is described by thermodynamics
and hydrodynamics, where we identify the expectation value in (\ref{eq:dictionary})
with the macroscopic (classical) mechanical quantity $O_{\phi}(x)$.

Two examples are of special interests. One is the case that $\phi$
is taken to be the metric $g_{\mu\nu}$, where $\bar{\phi}$ is just
the induced metric $\bar{g}_{ab}$ on the boundary. Then the Minkowskian
version of (\ref{eq:dictionary}) tells us that the stress-energy
tensor of the boundary system is given by the Brown-York tensor (see
(\ref{eq:Brown-York}) for the explicit form)
\[
t_{ab}(x)=\frac{2}{\sqrt{-g}}\frac{\delta I_{\mathrm{bulk}}[\bar{g}]}{\delta\bar{g}_{ab}(x)},
\]
where the bulk action is taken to be the standard Einstein-Hilbert
action plus the Gibbons-Hawking term. The other is the case that $\phi$
is taken to be the electromagnetic potential $A_{\mu}$. Similarly,
the dictionary is that the electric current of the boundary system
is given by
\begin{equation}
j^{a}(x)=\frac{1}{\sqrt{-g}}\frac{\delta I_{\mathrm{bulk}}[\bar{A}]}{\delta\bar{A}_{a}(x)}=-n_{\mu}F^{\mu a},\label{eq:electric}
\end{equation}
where the bulk action is just the Maxwell one in addition to the gravitational
part. In this section, we first explore the macroscopic aspects of
the general bulk/boundary correspondence in the equilibrium case,
based on the above holographic dictionary.

\subsection{Thermodynamics dual to the RN bulk space-time}

We consider the RN black hole
\begin{eqnarray}
ds_{d+1}^{2} & = & \frac{dr^{2}}{f(r)}-f(r)dt^{2}+r^{2}d\Omega_{d-1}^{(k)2},\nonumber \\
f(r) & = & k+\frac{r^{2}}{\ell^{2}}-\frac{2M}{r^{d-2}}+\frac{Q^{2}}{r^{2d-4}},\nonumber \\
d\Omega_{d-1}^{(k)2} & = & \hat{g}_{ij}^{(k)}(x)dx^{i}dx^{j},\nonumber \\
A & = & \sqrt{\frac{d-1}{8\pi(d-2)G}}\frac{Q}{r^{d-2}}dt,\label{eq:metric}
\end{eqnarray}
with negative cosmological constant%
\footnote{The case with positive cosmological constant can also be included
by formally allowing $\ell^{2}<0$.%
} in the Einstein-Maxwell theory as our bulk space-time (in equilibrium).
Here $M$ is the mass parameter, $Q$ the charge parameter of the
black hole, and $\hat{g}_{ij}^{(k)}(x)$ the metric on the ``unit''
sphere, plane or hyperbola for $k$ equal to $1$, $0$ or $-1$,
respectively, where in the planar or hyperbolic case some standard
compactification is assumed. The boundary is the hypersurface $r=r_{c}$
outside the horizon, with an induced metric
\begin{equation}
ds_{d}^{2}=-f_{c}dt^{2}+r_{c}^{2}d\Omega_{d-1}^{(k)2},\qquad f_{c}:=f(r_{c}).\label{eq:boundary}
\end{equation}
Due to static nature (with time-like Killing vector $\partial_{t}$)
of both the bulk space-time and the boundary, and maximum symmetry
on a time slice of the boundary, the boundary system is obviously
in equilibrium. From the identification (\ref{eq:Euclidean}) of the
Euclidean partition function (see the next subsection for detailed
discussions), an argument of conical singularity leads to the conclusion
that the entropy and temperature of the boundary system are equal
to the Bekestein-Hawking entropy
\[
S=\frac{\Omega_{d-1}^{(k)}r_{h}^{d-1}}{4G}
\]
and local Hawking temperature
\begin{equation}
T=\frac{T_{H}}{\sqrt{f_{c}}}=\frac{f_{h}^{\prime}}{4\pi\sqrt{f_{c}}},\qquad f_{h}^{\prime}:=f^{\prime}(r_{h})\label{eq:local_Hawking}
\end{equation}
of the bulk black hole. Here $\Omega_{d-1}^{(k)}$ is the volume of
the ``unit'' sphere, plane or hyperbola, and $r_{h}$ the radius
of the outer horizon satisfying $f(r_{h})=0$. Due to the bulk/boundary
dictionary, the stress-energy tensor of the boundary system is given
by the Brown-York tensor
\begin{equation}
t_{ab}=\frac{1}{8\pi G}(Kg_{ab}-K_{ab}-Cg_{ab}),\qquad K:=K_{ab}g^{ab}\label{eq:Brown-York}
\end{equation}
on the boundary with $K_{ab}$ its extrinsic curvature and $C$ some
constant, which can be easily shown to have a form of ideal fluid
\[
t_{ab}=\epsilon u_{a}u_{b}+p(u_{a}u_{b}+g_{ab})
\]
with the velocity $u_{a}=(-\sqrt{f_{c}},0,\cdots,0)$, the energy
density
\begin{equation}
\epsilon=-\frac{d-1}{8\pi G}\frac{\sqrt{f_{c}}}{r_{c}}+C,\label{eq:energy_density}
\end{equation}
and the pressure
\[
p=\frac{d-2}{8\pi G}\frac{\sqrt{f_{c}}}{r_{c}}+\frac{1}{16\pi G}\frac{f_{c}^{\prime}}{\sqrt{f_{c}}}-C.
\]
As well, the electric current (\ref{eq:electric}) of the boundary
system is
\begin{equation}
j^{a}=-n_{\mu}F^{\mu a}(r_{c})=(-\sqrt{\frac{(d-1)(d-2)}{8\pi Gf_{c}}}\frac{Q}{r_{c}^{d-1}},0,\cdots,0).\label{eq:current}
\end{equation}
Since the volume of the boundary system is
\[
V=\Omega_{d-1}^{(k)}r_{c}^{d-1},
\]
the energy density (\ref{eq:energy_density}) gives the total energy
\[
E=\Omega_{d-1}^{(k)}(-\frac{d-1}{8\pi G}\sqrt{f_{c}}r_{c}^{d-2}+Cr_{c}^{d-1}),
\]
while the electric current (\ref{eq:current}) gives the total charge
\[
\Omega_{d-1}^{(k)}\sqrt{\frac{(d-1)(d-2)}{8\pi G}}Q
\]
that coincides with the physical charge of the black hole. The proportion
coefficient here is not essential, so we will take $Q$ as the total
charge in the following discussion.

As a consistency check, if expressing $E$ as a function of $(S,V,Q)$,
one can verify
\[
\frac{\partial E}{\partial S}=T,\qquad\frac{\partial E}{\partial V}=-p.
\]
Furthermore, one can obtain
\begin{equation}
\mu=\frac{\partial E}{\partial Q}=-\frac{d-1}{8\pi G}\frac{\Omega_{d-1}^{(k)}Q}{\sqrt{f_{c}}}(\frac{1}{r_{c}^{d-2}}-\frac{1}{r_{h}^{d-2}}),\label{eq:chemical}
\end{equation}
which is proportional to the difference of electric potential between
the horizon and the holographic screen, and is the appropriate generalization
of the familiar chemical potential in AdS/CFT ($r_{c}\to\infty$).
As a consistency check, we will show shortly that the chemical potential
(\ref{eq:chemical}) gives the correct Einstein relation on the holographic
screen $r=r_{c}$. Thus, we see that the first law
\begin{equation}
dE+pdV=TdS+\mu dQ\label{eq:1st_law}
\end{equation}
of thermodynamics holds for the boundary system. In the plane symmetric
case ($k=0$), a further relation
\[
E+pV=TS+\mu Q
\]
holds as the Gibbs-Duhem relation, as one may expect from extensibility
arguments (see the next subsection). In this case, it is convenient
to express the thermodynamic relations in terms of densities of the
extensive quantities as \cite{NTWL}
\begin{eqnarray}
\epsilon+p & = & Ts+\mu\rho,\nonumber \\
d\epsilon & = & Tds+\mu d\rho,\label{eq:density}
\end{eqnarray}
where $s=\frac{S}{V}$ is the entropy density and $\rho=\frac{Q}{V}$
the charge density.

Now we show that the chemical potential (\ref{eq:chemical}) is consistent
with the Einstein relation
\begin{equation}
\sigma=\Xi D,\label{eq:Einstein}
\end{equation}
where $\sigma$ is the electric conductivity, $\Xi$ the susceptibility,
and $D$ the diffusion constant. Following the corresponding discussion
\cite{membrane} in the standard AdS/CFT, we work at the linear order
of $\rho$ in $\mu$, which is actually the limit of small $Q$. In
this case, we have $\rho=\Xi\mu$, so the susceptibility
\[
\Xi=\frac{\rho}{\mu}=\frac{\frac{(d-1)(d-2)}{8\pi G}\frac{Q}{r_{c}^{d-1}}}{-\frac{d-1}{8\pi G}\frac{Q}{\sqrt{f_{c}}}(\frac{1}{r_{c}^{d-2}}-\frac{1}{r_{h}^{d-2}})}=(d-2)\frac{\sqrt{f_{c}}}{r_{c}^{d-1}}(\frac{1}{r_{h}^{d-2}}-\frac{1}{r_{c}^{d-2}})^{-1}.
\]
The conductivity $\sigma$ and the diffusion constant $D$ have been
computed at the finite holographic screen \cite{membrane,Strominger,Zhou}.%
\footnote{Note that in Ref.\cite{membrane,Zhou}, the electric current at a
finite cutoff surface differs from our definition (\ref{eq:electric})
by a $\sqrt{-\bar{g}}$ factor, so does other quantities related to
conjugate momenta. Our formalism is close to Ref.\cite{Strominger},
which treats the holographic screen as an effective physical system,
so such quantities should be more suitably defined as intrinsic tensor
(vector, scalar) fields on the screen. Correspondingly, although the
famous ratio $\frac{\eta}{s}$ with $\eta$ the shear viscosity does
not run with $r_{c}$ in both formalism, $\eta$ and $s$ independently
do not run with $r_{c}$ in the formalism of Ref.\cite{membrane,Zhou},
while they do run with $r_{c}$ in our formalism. {See also the related discussion in Sec.\ref{sec:conformal}.}%
} In our notation and convention, the results are
\begin{eqnarray*}
\sigma & = & \frac{r_{h}^{d-3}}{r_{c}^{d-1}},\\
D & = & \frac{r_{h}^{d-3}}{(d-2)\sqrt{f_{c}}}(\frac{1}{r_{h}^{d-2}}-\frac{1}{r_{c}^{d-2}}),
\end{eqnarray*}
so it is obvious that the Einstein relation (\ref{eq:Einstein}) holds.

\subsection{The general thermodynamics by Hamilton-Jacobi-like analysis}

For more general gravitational theories with various matter content,
the dual thermodynamic relation similar to (\ref{eq:1st_law}) can
be obtained through a Hamilton-Jacobi-like analysis, which we present
here. There are several types of ensembles that we can choose. They
are related to one another by Legendre transformations, in the thermodynamic
limit. For the system with charges, such as that dual to the RN black
hole (\ref{eq:metric}), the most often used ensemble in AdS/CMT or
fluid/gravity correspondence is the grand-canonical ensemble, so we
will take this ensemble as our starting point. {Other ensembles can be discussed similarly.}
First of all, at finite temperature the holographic principle (\ref{eq:Euclidean})
is naturally extended to
\begin{equation}
\begin{array}{ccc}
\mbox{black-hole Euclidean partition function} & = & \mbox{(grand-)canonical partition function},\\
\mbox{(bulk)} &  & \mbox{(boundary)}
\end{array}\label{eq:partition}
\end{equation}
where for the grand-canonical case the black-hole Euclidean partition
function is evaluated under the boundary condition of fixed chemical
potential (\ref{eq:chemical}), instead of fixed charge $Q$ for the
canonical case. The black-hole Euclidean partition functions (or more
precisely the logarithm of them) under different sets of boundary
conditions are also related to one another by Legendre transformations,
in the classical limit of gravity. Here we see the classical limit/thermodynamic
limit correspondence in the general (Euclidean) bulk/boundary holography,
as already indicated in the standard AdS/CFT case.%
\footnote{{In AdS/CFT, the ``thermodynamic limit'' here manifests
itself more familiarly as a large $N$ limit or a large central charge
limit, which is apparently different from its usual meaning.}%
} In fact, using the Hamilton-Jacobi-like analysis and insisting on
the micro-canonical ensemble, Brown and York obtain the first law-like
relation from the purely gravitational point of view \cite{BY2}.
But we will present a simpler argument in the context of holography,
which in the same time clearly shows the relation of different ensembles.

A natural requirement for the spatial section of our holographic setup
is homogeneity and isotropy, since it is hard to define equilibrium
otherwise. Recall that the central quantity of the grand-canonical
ensemble is the grand potential $\Omega$, as a function of the temperature
$T$ (or the inverse temperature $\beta$), the chemical potential
$\mu$ and the volume $V$. Under the classical limit on the bulk
side and the thermodynamic limit on the boundary side, the holographic
principle (\ref{eq:partition}) becomes
\begin{equation}
\exp[-I_{{\rm bulk}}(\beta,\mu,V)]=\exp(\beta\Omega),\label{eq:grand_principle}
\end{equation}
where in the homogeneous and isotropic case the on-shell (Euclidean)
action $I_{{\rm bulk}}$ is evaluated under the boundary condition
of fixed $\beta$, $\mu$ and $V$. Here we have suppressed any other
possible fields in the theory, such as the scalar field in the holographic
superconductor/superfluid models, which are easily included in this
discussion. Note that from the holographic point of view, $\beta$
is just the periodicity of the Euclidean time, as a Killing vector
field of the bulk space-time, measured by the proper time on the boundary,
which is determined by the condition of regularity, i.e. having no
conical defect, at the (Euclidean) horizon.

Now we give two equivalent ways to vary $I_{{\rm bulk}}$ and then
compare the results. The first way is to vary $I_{{\rm bulk}}$ with
respect to $\beta$, $\mu$ and $\bar{g}_{ii}$ (the diagonal spatial
component of $\bar{g}_{ab}$). Since $I_{{\rm bulk}}$ is on shell,
similar to the standard Hamilton-Jacobi equation, variation of $I_{{\rm bulk}}$
with respect to $\mu=\bar{A}_{\tau}$ and $\bar{g}_{ii}$ while keeping
$\beta$ and the other quantities fixed gives
\begin{equation}
\delta I_{{\rm bulk}}=\rho\beta V\delta\mu+p\beta\delta V.\label{eq:mu_V}
\end{equation}
Note that the triviality of the Wilson loop of $A_{\mu}$ contracted
at the Euclidean horizon requires $A_{\tau}=0$ there automatically.
On the other hand, variation of $I_{{\rm bulk}}$ with respect to
$\beta$ gives \cite{conical}
\begin{equation}
\delta I_{{\rm bulk}}=\frac{I_{{\rm bulk}}}{\beta}\delta\beta-\frac{S}{\beta}\delta\beta,\label{eq:beta}
\end{equation}
where the last term can be viewed as coming from the contribution
of the conical singularity when $\beta$ is perturbed away from the
original periodicity of the Euclidean time. (See \cite{Maldacena_entropy}
for more detailed and rigorous discussions about subtleties that may
arise here.) Combining (\ref{eq:grand_principle}), (\ref{eq:mu_V})
and (\ref{eq:beta}), we obtain
\begin{equation}
d(\beta\Omega)=\Omega d\beta+\frac{S}{\beta}d\beta-\rho\beta Vd\mu-p\beta dV,\label{eq:variation}
\end{equation}
which is just the thermodynamic relation
\[
\beta d\Omega=\frac{S}{\beta}d\beta-\rho\beta Vd\mu-p\beta dV,
\]
or more conveniently
\begin{equation}
d\Omega=-SdT-Qd\mu-pdV.\label{eq:grand}
\end{equation}

The second way is to vary $I_{{\rm bulk}}$ with respect to $\bar{g}_{\tau\tau}$,
$\mu$ and $\bar{g}_{ii}$. Similar to (\ref{eq:variation}) in the
first way, we obtain
\begin{eqnarray*}
d(\beta\Omega) & = & \epsilon Vd\beta-\rho Vd(\beta\mu)-p\beta dV\\
 & = & (E-\mu Q)d\beta-\rho\beta Vd\mu-p\beta dV.
\end{eqnarray*}
Comparing the above equation with (\ref{eq:variation}), we see
\[
\Omega=E-TS-\mu Q,
\]
which together with (\ref{eq:grand}) shows that $\Omega$ is indeed
related to the energy $E$ by Legendre transformations. Then, the
standard first law
\[
dE+pdV=TdS+\mu dQ
\]
of thermodynamics follows immediately.

In the case with planar symmetry, there is another important relation.
Similar to the extensibility arguments in the ordinary thermodynamics
in textbooks, we have
\[
\Omega(T,\mu,\lambda V)=\lambda\Omega(T,\mu,V)
\]
with an arbitrary scaling parameter $\lambda$, since $V$ is the
only extensive quantity in the arguments of $\Omega$, and there is
no extra independent scale in the system. That gives
\[
\Omega=(\frac{\partial\Omega}{\partial V})_{T,\mu}V=-pV,
\]
and then the Gibbs-Duhem relation
\begin{equation}
E+pV=TS+\mu Q,\label{eq:Gibbs}
\end{equation}
which we have seen in the $\varepsilon=0$ case in the last subsection.
Sometimes this relation is also expressed as the differential form
\[
Vdp-SdT-Qd\mu=0.
\]

In the other cases (i.e. with spherical or hyperbolic symmetry), it
is expected that the usual Gibbs-Duhem relation (\ref{eq:Gibbs})
will no longer hold. Actually, one can easily show from the expressions
in the preceding subsection that $E+pV\ne TS+\mu Q$ if $k\ne0$ for
the RN black hole (\ref{eq:metric}). However, it turns out that a
peculiar Gibbs-Duhem-like relation still holds, at least formally.
In fact, let $k=\varepsilon^{\frac{2}{d-1}}$ in (\ref{eq:metric}),
one will obtain the conjugate quantity $\varsigma$ of $\varepsilon$
as
\begin{equation}
\varsigma=(\frac{\partial E}{\partial\varepsilon})_{S,Q,V}=-\frac{\Omega_{d-1}}{8\pi G}\frac{r_{c}^{d-2}-r_{h}^{d-2}}{\sqrt{f_{c}}}\varepsilon^{\frac{3-d}{d-1}}.\label{eq:sigma}
\end{equation}
Note that the volume $\Omega_{d-1}$ of the ``unit'' sphere, plane
or hyperbola can be arbitrarily dependent on $k$ in the discussion
of the preceding subsection, but here we assume that this volume is
a constant independent of $k$, so that the total volume $V$ of the
boundary system can be simply identified with $r_{c}^{d-1}$ when
varying $k$. Then one can easily check that the Gibbs-Duhem-like
relation
\begin{equation}
E+pV=TS+\mu Q+\varsigma\varepsilon\label{eq:Gibbs-like}
\end{equation}
holds. But how can one understand the physical or geometric meaning
of $\varepsilon$ (and $\varsigma$)? The $d=3$ case, where the bulk
is of four dimensions and $k=\varepsilon$, is the simplest one to
illustrate the meaning of $\varepsilon$. In this case $f(r)=\varepsilon+\frac{r^{2}}{\ell^{2}}-\frac{2M}{r}+\frac{Q^{2}}{r^{2}}$,
it is easy to see that the transformation
\[
\varepsilon\to\lambda^{2}\varepsilon,\qquad r\to\lambda r,\qquad M\to\lambda M,\qquad Q\to\lambda Q
\]
\[
t\to\lambda^{-1}t,\qquad d\Omega_{2}^{2}\to\lambda^{-2}d\Omega_{2}^{2}
\]
leaves the configuration (\ref{eq:metric}) invariant. Using the
above transformation with $\lambda=(-\varepsilon)^{-1/2}$, one can
transform an $\varepsilon\ne-1$ solution to an $\varepsilon=-1$ one.
However, this transformation has an additional consequence
$\Omega_{2}\to-\varepsilon\Omega_{2}$. If $\varepsilon$ is a
negative integer, this multiplies the volume of the ``unit
hyperbola'' by an integer. Recalling that every two dimensional
compact surface with constant negative curvature is the original
hyperbolic space (Poincar\'e upper-half plane) modulo some discrete
group (see e.g. \cite{Polchinski}), one recognize $\varepsilon$ as
(half of) the Euler number of the spatial section of the black hole
horizon. This interpretation remains valid for $\varepsilon=0,1$,
but becomes obscure for $\varepsilon>1$ or $\varepsilon$ not
integer, in which case the interpretation of $\varepsilon$ as a
Euler number is only formal. Thus, at least formally, one sees that
including the \emph{topological charge} $\varepsilon$, as well as
the (gauge) charge $Q$, as thermodynamic quantities for the case
without planar symmetry, the Gibbs-Duhem-like relation
(\ref{eq:Gibbs-like}) can be obtained. Now we consider the $d\ne3$
case. If $d$ is even, then the spatial section of the horizon is of
odd dimensions and has no well-defined Euler number. In the same
time, $\varepsilon=k^{\frac{d-1}{2}}$ is complex for negative $k$,
so (\ref{eq:Gibbs-like}) is very formal in this case. If $d$ is odd,
the Euler density of the section scales as $R^{\frac{d-1}{2}}$ with
$R$ the curvature tensor of the section, the similar discussion as
in the $d=3$ case above also leads to the conclusion that
$\varepsilon$ can be viewed as (half of) the Euler
number of the section.%
\footnote{Note that if we let $k=\varepsilon^{\alpha}$ with $\alpha\ne\frac{2}{d-1}$,
then the relation (\ref{eq:Gibbs-like}) cannot hold, so the validity
of the Gibbs-Duhem-like relation can be viewed as equivalent to the
fact that $\varepsilon$ is the topological charge.%
} Strictly speaking, the topological charge $\varepsilon$ should be
an integer, as well as the charge $Q$ should be quantized, which
makes the corresponding first law-like relation
\begin{equation}
dE+pdV=TdS+\mu dQ+\varsigma d\varepsilon\label{eq:1st_law_plus}
\end{equation}
less interesting. However, when the above relation is expressed in
terms of densities:
\[
d\epsilon=Tds+\mu d\rho+\varsigma de
\]
with the topological charge density $e=\frac{\varepsilon}{V}$, just
as (\ref{eq:density}), the quantization of $\varepsilon$ and $Q$
can be smoothed out by a large volume $V$ of the boundary system.\footnote{{If
considering the $d$-dimensional spatial section of the bulk black hole,
the density $e$ is just the local flux of a recently proposed ``Euler current'' \cite{GRS}
across the $(d-1)$-dimensional spatial section of the hypersurface $r=r_c$.
However, further consequences of this fact still need to be investigated.}}
The expression (\ref{eq:sigma}) of $\varsigma$ also looks like that
of the chemical potential (\ref{eq:chemical}), but it is not clear
whether $\varsigma$ can be understood as the difference of some kind
of potential. Furthermore, one may expect that the topological charge
$\varepsilon$ will also play some role in the ordinary thermodynamics
of black holes. We refer the interested readers to Appendix \ref{sec:thermodynamics}.

\section{Entropy production on the holographic screen and its equality with
the increase of entropy in the bulk}

If the boundary system is perturbed by some sort of external sources,
various transport processes occur {as the system relaxes back
towards equilibrium}, which causes entropy production. From the bulk point
of view, the ingoing boundary condition at the future horizon implies
that the (material or gravitational) perturbations at the boundary
should propagate to the black hole and be absorbed, which causes increase
of the area of the black hole horizon. Based on the equilibrium configuration
we have discussed above, there are three kinds of transport processes
that we can consider, i.e. heat conduction, viscosity of fluid and
charge conduction. The heat conduction is energy transportation, caused
by temperature inhomogeneity. The viscosity of fluid is momentum transportation,
caused by velocity inhomogeneity (shear and expansion, more precisely).
The charge conduction is caused by external electric field or inhomogeneity
of chemical potential. Then a slight generalization of the textbook
discussion (see, e.g. Ref.\cite{Reichl}) to the case allowing a curved
space gives the total entropy production rate%
\footnote{Note that two kinds of independent viscous processes, i.e. shear viscosity
and bulk viscosity, present here, whose contributions to the entropy
production rate are written in a uniform way. The shear viscosity
is the transport process for the tangent component of momentum, and
the bulk viscosity is the transport process for the normal component
of momentum.%
}
\[
\Sigma=\mathbf{j}_{q}\cdot\mathbf{D}\frac{1}{T}-\frac{1}{T}\Pi:\mathbf{D}\mathbf{u}+\frac{1}{T}\mathbf{j}\cdot\mathbf{E}=j_{q}^{i}D_{i}\frac{1}{T}-\frac{1}{T}\Pi^{ij}\sigma_{ij}+\frac{1}{T}j^{i}E_{i},
\]
where $\mathbf{j}_{q}$ is the heat current, $\Pi^{ij}$ the dissipative
part of the stress-energy tensor, $\sigma_{ij}=D_{(i}u_{j)}\equiv\frac{1}{2}(D_{i}u_{j}+D_{j}u_{i})$
{the combination of shear tensor and expansion rate}, $\mathbf{j}$
the electric current, $\mathbf{E}$ the electric field, and we have
assumed a homogeneous chemical potential. {With a slight abuse of terminology, we also call $\sigma_{ij}$
the shear tensor for the sake of briefness, but keep in mind that its traceless part is the genuine
shear tensor while its trace part is the expansion rate.} In fact, the physical laws
of transportation tell us that the transport current $(j_{q}^{i},\Pi^{ij},j^{i},\cdots)$
is proportional (in the linear regime) to the driving force $(D_{i}\frac{1}{T},-\frac{1}{T}\sigma_{ij},\frac{1}{T}E_{i},\cdots)$,
while the entropy production rate is just their product. In the holographic
context, this proportion factor (matrix), i.e. the transport coefficients,
is determined by imposing the ingoing boundary condition at the horizon
and then solving the bulk equations of motion (see e.g. Ref.\cite{Hartnoll,Herzog}
for the traditional AdS/CFT case and Ref.\cite{Strominger,Zhou,GLTW}
for the ``finite cutoff'' case). However, we do not need the precise
values of them here. {Note that on the bulk side we only consider
classical gravitational theory with classical matter fields.}

It is well known in AdS/CFT that the temperature inhomogeneity and
shear can both be realized by gravitational perturbations, at least
for some special configurations (see, e.g. \cite{Hartnoll,Herzog}
and \cite{S S}, respectively). Now we generalize the analyses to
arbitrary (but small) temperature perturbation and shear field on
the holographic screen $r=r_{c}$. For the sake of simplification,
we do a constant rescaling of $t$ such that the metric (\ref{eq:boundary})
on the holographic screen becomes
\begin{equation}
ds_{d}^{2}=-dt^{2}+\cdots.\label{eq:rescale_t}
\end{equation}
The temperature perturbation can be introduced by the metric perturbation
\[
ds_{d}^{2}\to ds_{d}^{2}+2h_{ti}dtdx^{i},
\]
generalizing the discussion in \cite{Hartnoll,Herzog}, as
\[
D_{i}\frac{1}{T}=\frac{1}{T}\partial_{t}h_{ti},
\]
{which can be briefly argued as follows.
First, we perturb the time-time component of the metric by $h_{tt}$ before turning on
an off-diagonal metric perturbation $h_{ti}$. Recall that the inverse temperature
$\beta=1/T$ is the period of the Euclidean time measured under the proper time units,
so the metric perturbation $h_{tt}$ induces the temperature perturbation
\[
\frac{T_0^2}{T^2}=1-h_{tt}
\]
with $T_0$ the constant equilibrium temperature, or in other words
\[
\partial_i h_{tt}=2\frac{\partial_i T}{T}
\]
at leading (linear) order. Next, we make an infinitesimal coordinate transformation
(diffeomorphism) to turn off $h_{tt}$ and exhibit the temperature gradient
by the off-diagonal metric perturbation $h_{ti}$. Using $\mathcal{L}_\xi g_{ab}=D_a\xi_b+D_b\xi_a$, it is easy to check
that the diffeomorphism induced by the vector field $\xi$ satisfying
\[
\partial_t\partial_i\xi_t=-\frac{\partial_i T}{T},\qquad\xi_i=0
\]
can do the task, which turns on an off-diagonal metric perturbation
\[
\partial_{t}h_{ti}=-\frac{\partial_i T}{T}
\]
and then completes our argument. Note that all the covariant derivatives $D_a$ here have been
replaced by the partial derivatives $\partial_a$, since for the induced metric (\ref{eq:boundary})
the Christoffel symbol vanishes when any of its indices is $t$.}

Then, we insist on the frame $u^{a}=(1,0,\cdots,0)$, while turn on
the shear (and expansion, as always) by a metric perturbation
\[
ds_{d}^{2}\to ds_{d}^{2}+h_{ij}dx^{i}dx^{j}.
\]
In this case the shear tensor reads
\[
\sigma_{ij}=\partial_{(i}u_{j)}-\gamma_{ij}^{a}u_{a}=\frac{1}{2}(\partial_{i}\tilde{g}_{tj}+\partial_{j}\tilde{g}_{ti})-\gamma_{tij}=\frac{1}{2}\partial_{t}h_{ij}
\]
with $\tilde{g}_{ab}=g_{ab}+h_{ab}$ the perturbed metric and $\gamma_{bc}^{a}$
the corresponding (perturbed) Christoffel symbol. The heat current
$j_{q}^{i}$ is just the energy current $-t^{ti}$ (that vanishes
in equilibrium),%
\footnote{Note that in the case with cross-transportation, there is a mixing
between the heat current and the charge current, as well as that between
the temperature gradient and the potential gradient, just as shown
in Ref.\cite{Hartnoll,Herzog}. However, this complication does not
ruin our discussion on the entropy production rate, since that rate
is the very bilinear of conjugate quantities that is invariant under
such mixing.%
} while $\Pi^{ij}$ is just the (first order) perturbation of $t^{ij}$.
So we have the total entropy production rate
\begin{equation}
\Sigma=-\frac{1}{2T}t^{(1)ab}\partial_{t}h_{ab}+\frac{1}{T}j^{i}E_{i},\label{eq:production}
\end{equation}
where the contribution from the charge conduction is simply realized
by electromagnetic perturbations. Our central task in this section
is to check whether (\ref{eq:production}) matches the black hole
side.

\subsection{The case without cross-transportation}

For clarity, we first assume that the equilibrium background is uncharged,
i.e. $Q=0$. Since in this case the gravitational perturbation and
electromagnetic perturbation are decoupled from each other, it turns
out that the first two kinds of transport processes and the charge
conduction are decoupled, which allows us to discuss them separately.
First, we consider gravitational perturbations, realizing the heat
conduction and viscosity of {the dual} fluid. For convenience, we rescale the
$r$ coordinate such that
\begin{equation}
ds_{d+1}^{2}=dr^{2}+g(r)dt^{2}+\cdots,\label{eq:rescale_r}
\end{equation}
{setting} $g(r_{c})=-1$ to guarantee (\ref{eq:rescale_t}). Now we introduce
the gravitational perturbation with gauge
\begin{equation}
g_{\mu\nu}\to g_{\mu\nu}+h_{\mu\nu},\qquad h_{r\mu}=0\label{eq:perturbation}
\end{equation}
in the bulk, while in addition the time-time component $h_{tt}$ vanishes
on the holographic screen. The above metric implies the extrinsic
curvature
\[
K_{ab}=\frac{1}{2}\mathcal{L}_{n}g_{ab}=\frac{1}{2}\partial_{r}g_{ab}\rightarrow\frac{1}{2}\partial_{r}(g_{ab}+h_{ab})
\]
for any hypersurface of constant $r$, with $n$ its unit normal,
even after perturbation (\ref{eq:perturbation}). On the boundary
side, since the background Brown-York tensor $t_{ab}$ has no off-diagonal
elements, the entropy production rate (\ref{eq:production}) is obviously
of order $\mathcal{O}(h_{ab}^{2})$. To leading order, the entropy
production rate (\ref{eq:production}) without the electromagnetic
part is
\begin{eqnarray}
\Sigma & = & -\frac{1}{2T}(t_{cd}^{(1)}g^{ca}g^{db}-t_{cd}^{(0)}h^{ca}g^{db}-t_{cd}^{(0)}g^{ca}h^{db})\partial_{t}h_{ab}\nonumber \\
 & = & -\frac{1}{2T}(t_{ab}^{(1)}\partial_{t}h^{ab}-2t_{c}^{(0)b}h^{ca}\partial_{t}h_{ab})\nonumber \\
 & = & -\frac{1}{16\pi GT}(K^{(1)}g_{ab}+K^{(0)}h_{ab}-K_{ab}^{(1)}-2K^{(0)}h_{ab}+2K_{b}^{(0)c}h_{ca})\partial_{t}h^{ab}\nonumber \\
 & = & -\frac{1}{16\pi GT}(\frac{1}{2}\partial_{r}h\partial_{t}h-K^{(0)}h_{ab}\partial_{t}h^{ab}-\frac{1}{2}\partial_{r}h_{ab}\partial_{t}h^{ab}+2K_{b}^{(0)c}h_{ca}\partial_{t}h^{ab}),\label{eq:rate}
\end{eqnarray}
where we have defined $h\equiv g^{ab}h_{ab}$, and the indices are
lowered (or raised) with the background metric $g_{ab}$. Here the
superscript $(0)$ and $(1)$ denote the background (equilibrium)
quantities and the first order variations induced by the gravitational
perturbation $h_{ab}$, respectively. Particularly, $K^{(1)}$ means
the first order perturbation of $K=K_{ab}g^{ab}$, which can be written as
\[
K^{(1)}=K_{ab}^{(1)}g^{ab}-K_{ab}^{(0)}h^{ab}
=\frac{1}{2}g^{ab}\partial_{r}h_{ab}-\frac{1}{2}h^{ab}\partial_{r}g_{ab}=\frac{1}{2}\partial_r h.
\]
On the bulk side, the physical
picture is that the gravitational wave caused by the boundary perturbation
propagates to the black hole, which will be absorbed and render the
horizon area to increase. From the theory of gravitational waves (see,
e.g. \cite{gravity_wave}), we know that the effective stress-energy
tensor $T_{\mu\nu}$ of the wave is just $-\frac{1}{8\pi G}$ times
$G_{\mu\nu}^{(1,1)}$, the second order contribution of $h_{\mu\nu}$
to the Einstein tensor, which satisfies
\[
\nabla_{\mu}T^{\mu\nu}=0
\]
to order $\mathcal{O}(h_{ab}^{2})$ with respect to the background
covariant derivative $\nabla_{\mu}$. Since $\xi=\partial_{t}$ is
Killing, we have the conservation law
\begin{equation}
\nabla_{\mu}(T_{\nu}^{\mu}\xi^{\nu})=0\label{eq:conserve}
\end{equation}
of the current $T_{\nu}^{\mu}\xi^{\nu}=T_{t}^{\mu}$. Physically,
our whole setup is in a static (equilibrium) state before the perturbation
is turned on at some time, then the perturbation on the boundary system
dissipates due to the transport processes, while the perturbation
in the bulk (and on the horizon) gradually fades away due to the black
hole absorption. Hence integrating the above equation over the perturbed
bulk region with the perturbed horizon as the inner boundary and using
the Gauss law, we end up with
\begin{equation}
\int_{H}T_{t}^{\mu}\lambda_{\mu}\tilde{\epsilon}_{[d]}=\int_{\mathrm{bdry}}T_{t}^{\mu}n_{\mu}\hat{\epsilon}_{[d]},\label{eq:correspond}
\end{equation}
where $H$ is the horizon and $\hat{\epsilon}_{[d]}=\sqrt{-\bar{g}}d^{d}x$.
Here $\lambda^{\mu}$ is tangent to the affinely parametrized null
geodesic generators of $H$, and $\tilde{\epsilon}_{[d]}$ satisfies
\[
\lambda\wedge\tilde{\epsilon}_{[d]}=-\epsilon_{[d+1]}
\]
with $\epsilon_{[d+1]}$ the standard volume element in the bulk.
The left hand side of (\ref{eq:correspond}) is just
the heat absorbed by the black hole \cite{Wald94,G W}{, which satisfies}
\begin{equation}
\int_{H}T_{t}^{\mu}\lambda_{\mu}\tilde{\epsilon}_{[d]}=T_{H}\delta
S\label{eq:first_law}.
\end{equation}
This will be shown in Appendix \ref{sec:Raychaudhuri}.

To evaluate the right hand side of (\ref{eq:correspond}), we should
know the explicit form of $G_{\mu\nu}^{(1,1)}$. From
\[
G_{\mu\nu}^{(1,1)}=R_{\mu\nu}^{(1,1)}-\frac{1}{2}(R^{(1,1)}g_{\mu\nu}+R^{(1)}h_{\mu\nu})
\]
and the Einstein equations at the zeroth and first orders, it is not
difficult to obtain
\[
G_{\mu\nu}^{(1,1)}=R_{\mu\nu}^{(1,1)}-\frac{1}{2}R_{\alpha\beta}^{(1,1)}g^{\alpha\beta}g_{\mu\nu}.
\]
The second order contribution of $h_{\mu\nu}$ to the Ricci tensor
$R_{\mu\nu}^{(1,1)}$ is given by \cite{MTW}
\begin{eqnarray*}
R_{\mu\nu}^{(1,1)} & = & \frac{1}{2}[\frac{1}{2}h_{\alpha\beta|\mu}h_{|\nu}^{\alpha\beta}+h^{\alpha\beta}(h_{\alpha\beta|\mu\nu}+h_{\mu\nu|\alpha\beta}-h_{\alpha\mu|\nu\beta}-h_{\alpha\nu|\mu\beta})+h_{\nu}^{\alpha|\beta}(h_{\alpha\mu|\beta}-h_{\beta\mu|\alpha})\\
 &  & -(h_{|\beta}^{\alpha\beta}-\frac{1}{2}h^{|\alpha})(h_{\alpha\mu|\nu}+h_{\alpha\nu|\mu}-h_{\mu\nu|\alpha})],
\end{eqnarray*}
where the indices are lowered (or raised) with the background metric
$g_{\mu\nu}$ and ``$|$'' denotes the background covariant derivative
$\nabla$. Some lengthy but straightforward calculation gives
\begin{eqnarray}
G_{rt}^{(1,1)}(r_{c}) & = & \frac{1}{2}[-\frac{1}{2}\partial_{r}h_{ab}D_{t}h^{ab}+2K_{a}^{(0)c}h_{cb}D_{t}h^{ab}-K_{t}^{(0)a}h^{bc}D_{a}h_{bc}\nonumber \\
 &  & +(K_{t}^{(1)a}-K_{tc}^{(0)}h^{ac})D_{a}h+D_{a}J^{a}],\label{eq:G_=00007Brt=00007D}
\end{eqnarray}
where $D_{a}$ is the background covariant derivative on the screen
and $J^{a}$ an order $\mathcal{O}(h_{ab}^{2})$ current. We do not
need the explicit form of $J^{a}$, for the divergence term $D_{a}J^{a}$
on the screen does not contribute to the right hand side of (\ref{eq:correspond})
in our case. Note that we also have the first two order ``momentum
constraints''
\begin{equation}
D_{a}t^{(0)ab}=\frac{1}{8\pi G}D_{a}(K^{(0)}g^{ab}-K^{(0)ab})=0,\label{eq:momentum_0th}
\end{equation}
\begin{equation}
D_{a}t_{b}^{(1)a}+D_{a}^{(1)}t_{b}^{(0)a}=D_{a}(K_{b}^{(1)a}-K_{bc}^{(0)}h^{ca}-K^{(1)}\delta_{b}^{a})+\gamma_{ac}^{(1)a}K_{b}^{(0)c}-\gamma_{ab}^{(1)c}K_{c}^{(0)a}=0.\label{eq:momentum_1st}
\end{equation}
where we have used the fact that the first order perturbation $D^{(1)}$
of the covariant derivative on the screen comes essentially from the
first order perturbation $\gamma^{(1)}$ of the corresponding Christoffel
symbol. For the latter, we have
\begin{equation}
\gamma_{ab}^{(1)c}=\frac{1}{2}g^{cd}(D_{a}h_{bd}+D_{b}h_{ad}-D_{d}h_{ab}),\qquad\gamma_{ac}^{(1)a}=\frac{1}{2}g^{ad}D_{c}h_{ad}=\frac{1}{2}D_{c}h,\label{eq:gamma_1st}
\end{equation}
which gives
\begin{equation}
D_{a}(K_{t}^{(1)a}-K_{tc}^{(0)}h^{ca}-K^{(1)}\delta_{t}^{a})+\frac{1}{2}K_{t}^{(0)c}D_{c}h-\frac{1}{2}K^{(0)ca}D_{t}h_{ac}=0\label{eq:momentum_1st^p}
\end{equation}
from the $b=t$ component of (\ref{eq:momentum_1st}). Since the background
extrinsic curvature $K_{ab}^{(0)}$ has no $K_{tj}^{(0)}$ (or $K_{it}^{(0)}$)
components and isotropy of the background space leads to $K_{ij}^{(0)}=\kappa g_{ij}$,
we know
\[
K^{(0)ca}D_{t}h_{ac}=\kappa D_{t}h,
\]
so (\ref{eq:momentum_1st^p}) multiplying by $h$ gives
\begin{equation}
(K_{t}^{(1)a}-K_{tc}^{(0)}h^{ca})D_{a}h\simeq K^{(1)}D_{t}h=\frac{1}{2}\partial_{r}h\partial_{t}h,\label{eq:KDh}
\end{equation}
where ``$\simeq$'' stands for equality up to divergence terms on
the screen. Substituting (\ref{eq:G_=00007Brt=00007D}) into (\ref{eq:correspond})
and using (\ref{eq:momentum_0th},\ref{eq:KDh}) when comparing with
(\ref{eq:rate}), we see from (\ref{eq:first_law}) that
\[
T_{H}\delta S=T\int_{\mathrm{bdry}}\Sigma\hat{\epsilon}_{[d]}.
\]
Upon identification $T_{H}=T$ due to our setting
(\ref{eq:rescale_t}),\footnote{{It is easy to incorporate the corresponding red-shift factor
in our discussion without setting (\ref{eq:rescale_t}), but we leave this generality to Section \ref{general}.}}
we conclude that the entropy increase on the bulk side and
the entropy production on the boundary side match exactly.

%Strictly speaking, the so-called physical process
%version (\ref{eq:1st_law}) of the first law of thermodynamics, which
%can be deduced from the Raychaudhuri equation, only applies to the
%case that $T_{\mu\nu}$ is a genuine stress-energy tensor of matter
%fields. In the above discussion, however, we have taken for granted
%that it also applies to the case that $T_{\mu\nu}$ is just an
%effective one of the gravitational wave. We will fill this gap in
%Appendix \ref{sec:Raychaudhuri}.

Next, we consider electromagnetic perturbations, which is much simpler.
For the electromagnetic wave, the physical picture is similar to the
gravitational case, except that the component of the stress-energy
tensor appearing in (\ref{eq:correspond}) is
\begin{equation}
T_{rt}(r_{c})=F_{r}^{\ i}(r_{c})F_{ti}(r_{c})=j^{i}E_{i},\label{eq:T_=00007Brt=00007D}
\end{equation}
where in the second equality we have used the holographic dictionary.
Then, combining (\ref{eq:production}) {(with the metric perturbation $h_{ab}$ turned off)},
(\ref{eq:correspond}), (\ref{eq:first_law}) and $T_{H}=T$, we obtain
\begin{equation}
\delta S=\int_{\mathrm{bdry}}\Sigma\hat{\epsilon}_{[d]}.\label{eq:entropy_increase}
\end{equation}
To sum up, for the uncharged background, we see perfect matching between
the entropy production from the above three kinds of transport processes
on the boundary and the entropy increase of the black hole in the
bulk.

\subsection{For more general gravitational theories}\label{general}

{We have shown by the above direct calculation the
consistency of the bulk entropy increase and the boundary entropy
production in the Einstein-Maxwell theory.} For more general
gravitational theories, similar calculations may be very difficult.
In this subsection, we give a more formal derivation of the same
result, which allows working in a more general class of
gravitational theories. For illustration purpose, now we do not
rescale $g_{tt}$ on the holographic screen to $-1$, so $u^{a}$
should be taken as $(1/\sqrt{f_{c}},0,\cdots,0)$. As well, we do not
transform $r$ to obtain a bulk metric of the form
(\ref{eq:rescale_r}) in this formal derivation. On the boundary
side, similar discussion leads to a total entropy production rate
\[
\Sigma=-\frac{1}{2\sqrt{f_{c}}T}t^{(1)ab}\partial_{t}h_{ab}+\frac{1}{T}j^{i}E_{i}
\]
instead of (\ref{eq:production}). Noting
\begin{eqnarray*}
D_{(t}u_{i)} & = & \partial_{(t}u_{i)}-\gamma_{ti}^{a}u_{a}=\frac{1}{2\sqrt{f_{c}}}(\partial_{t}\tilde{g}_{ti}+\partial_{i}\tilde{g}_{tt})-\gamma_{tti}=\frac{1}{2\sqrt{f_{c}}}\partial_{t}h_{ti},\\
D_{t}u_{t} & = & \partial_{t}u_{t}-\gamma_{tt}^{a}u_{a}=0,
\end{eqnarray*}
we can rewrite the above entropy production rate as
\begin{equation}
\Sigma=\frac{1}{T}j^{i}E_{i}-\frac{1}{T}t^{(1)ab}D_{a}u_{b}=\frac{1}{T}j^{i}E_{i}-\frac{1}{T}t^{(1)ab}(-\gamma_{ab}^{(1)c}u_{c}+D_{a}u_{b}^{(1)}),\label{eq:invariant}
\end{equation}
which is invariant under diffeomorphisms in the boundary system. In
fact,
\begin{equation}
D_{a}u_{b}^{(1)}=D_{a}(h_{bc}u^{c})=-\sqrt{f_{c}}D_{a}h_{b}^{t},
\end{equation}
and $\gamma_{ab}^{(1)c}$ is given by (\ref{eq:gamma_1st}).

We shall prove that the increase of black hole entropy in the bulk
is precisely equal to the aforementioned entropy production on the
holographic screen. As in the previous subsection, on the bulk side,
the basic idea for such a proof is to relate the holographic screen
to the bulk black hole horizon by the conserved current (\ref{eq:conserve})
with $\xi=\partial_{t}$. For the electromagnetic perturbation, the
conserved current is just generated by the stress-energy tensor
\begin{equation}
T^{\mu\nu}=\frac{2}{\sqrt{-g}}\frac{\delta I_{{\rm EM}}}{\delta g_{\mu\nu}}=F^{\mu\rho}F_{\ \rho}^{\nu}-\frac{1}{4}g^{\mu\nu}F_{\rho\sigma}F^{\rho\sigma}
\end{equation}
of the electromagnetic field, with its $rt$ component on the holographic
screen given by (\ref{eq:T_=00007Brt=00007D}). Then, integrating
the conserved current (\ref{eq:conserve}) and using the Gauss law
together with the first-law-like relation (\ref{eq:1st_law}) as in
the previous subsection, and noting
\[
\frac{1}{T}j^{i}E_{i}=\frac{1}{\sqrt{f_{c}}T}n_{\mu}F^{\mu i}(r_{c})F_{ti}(r_{c})
\]
by the holographic dictionary and the relation (\ref{eq:local_Hawking})
between the temperature $T$ on the screen and the black hole temperature
$T_{H}$ now, we see that (\ref{eq:entropy_increase}) holds for the
electromagnetic part of our correspondence, which is independent of
whatever gravitational theories we work in.

Next we consider the entropy production induced by the gravitational
perturbation on the boundary. As promised, we can work in a more general
class of gravitational theories, e.g. the general Lovelock gravity
\cite{Lovelock} as will be illustrated here. To proceed, let us firstly
expand the bulk equations of motion on the black hole background to
second order, i.e.,
\begin{eqnarray}
E^{\mu\nu}[g] & = & 0,\\
E^{(1)\mu\nu}[g,h] & = & 0,\label{dual1}\\
E^{(0,2)\mu\nu}[g,q] & = & -E^{(1,1)\mu\nu}[g,h]=:8\pi GT^{{\rm G}\mu\nu},\label{dual2}
\end{eqnarray}
where the metric is expanded as $g_{\mu\nu}+h_{\mu\nu}+q_{\mu\nu}$
with {$q_{\mu\nu}$ the second order perturbation and the indices are} raised or lowered by the background metric $g_{\mu\nu}$.
Furthermore, it follows from diffeomorphism invariance that the effective
gravitational energy-momentum tensor $T^{{\rm G}\mu\nu}$ is conserved
for the gravitational waves propagating on the background, i.e.
\[
\nabla_{\mu}T^{{\rm G}\mu\nu}=0,
\]
which, as before, gives rise to
\begin{equation}
\delta S=\frac{1}{T_{H}}\int_{H}T^{{\rm G}\mu\nu}\lambda_{\mu}\xi_{\nu}=\frac{1}{T_{H}}\int_{\mathrm{bdry}}T^{{\rm G}\mu\nu}n_{\mu}\xi_{\nu}.
\end{equation}

In Einstein's gravity, we know the Gauss-Codazzi equations
\[
D_{a}t^{ab}=-\frac{1}{8\pi G}G^{\mu b}n_{\mu}
\]
holding as geometric identities, which combines with the Einstein
equations to give the momentum constraints. The counterparts of these
constraints in the general Lovelock gravity can also be derived. The
action functional of the Lovelock gravity is
\begin{equation}
I=\int_{\Omega}L+\int_{\partial\Omega}B,\label{eq:surface}
\end{equation}
where $B$ is the surface term \cite{Myers,DBS} generalizing the
usual Gibbons-Hawking term for Einstein's gravity. By construction
of $B$, variation of the above action functional gives
\[
\delta I=\int_{\Omega}(-\frac{1}{2}E^{MN}\delta g_{MN})+8\pi G\int_{\partial\Omega}\frac{1}{2}t^{ab}\delta\bar{g}_{ab}
\]
with $t_{ab}$ the generalization of the Brown-York surface tensor,
which is holographically interpreted as the stress-energy tensor of
the boundary system. Suppose that the above metric variation is a
diffeomorphism induced by a vector field $\xi$ tangent to $\partial\Omega$,
the above equation becomes
\begin{eqnarray*}
\mathcal{L}_{\xi}I & = & \int_{\Omega}(-E^{MN}\nabla_{M}\xi_{N})+8\pi G\int_{\partial\Omega}t^{ab}D_{a}\xi_{b}\\
 & = & \int_{\partial\Omega}(-E^{Mb}n_{M}-8\pi GD_{a}t^{ab})\xi_{b},
\end{eqnarray*}
where we have used $\nabla_{M}E^{MN}=0$ (as a generalization of the
Bianchi identity) coming from the diffeomorphism invariance of $I$
without considering the boundary. The above expression vanishes due
to the diffeomorphism invariance of $I$, then the arbitrariness of
$\xi_{b}$ gives
\[
D_{a}t^{ab}=-\frac{1}{8\pi G}E^{\mu b}n_{\mu}.
\]
Further, the right hand side of the above equation vanishes due to
the equations of motion, which is just the momentum constraints.

So now the task boils down into whether one can express the above
flux across the holographic screen in terms of entropy production
on the screen, which can be achieved by a straightforward but lengthy
calculation as in the previous subsection. But here we would like
to present a shortcut towards the final result by taking advantage
of the dual role played by the gravitational waves. Namely, as demonstrated
in Eqs.(\ref{dual1}) and (\ref{dual2}), the gravitational waves,
albeit treated as sort of matter waves like light, are essentially
ripples in the fabric of space-time. Thus we can relate the aforementioned
flux to the quantities for the dual system on the holographic screen
by the corresponding momentum constraints, which, expanded to second
order, reads%
\footnote{Note that the perturbation of our boundary stress-energy tensor $t_{ab}$
is only induced by $h_{\mu\nu}$, not by $q_{\mu\nu}$. Here our focus is on the
tensor $T^{{\rm G}\mu\nu}$, which is just $\frac{-1}{8\pi G}E^{(1,1)\mu\nu}[g,h]$ by definition (\ref{dual2}).
Note that $D_{a}t^{ab}=-\frac{1}{8\pi G}E^{\mu b}n_{\mu}$ is a geometric identity,
which holds for any metric, so we take the metric in this identity to be $g_{\mu\nu}+h_{\mu\nu}$
(without $q_{\mu\nu}$) and do the following expansion. The relevance of the second order
perturbation $q_{\mu\nu}$ is reflected in the first equality of (\ref{dual2}).%
}
\begin{eqnarray}
D_{a}t^{(0)ab} & = & -\frac{1}{8\pi G}E^{\mu b}[g]n_{\mu}=0,\label{eq:constraint_0th}\\
D_{a}t^{(1)ab}+D_{a}^{(1)}t^{(0)ab} & = & -\frac{1}{8\pi G}E^{(1)\mu b}[g,h]n_{\mu}=0,\label{eq:constraint_1st}\\
D_{a}t^{(2)ab}+D_{a}^{(1)}t^{(1)ab}+D_{a}^{(2)}t^{(0)ab} & = & -\frac{1}{8\pi G}E^{(1,1)\mu b}[g,h]n_{\mu}=T^{{\rm G}\mu b}n_{\mu},
\end{eqnarray}
where $D_{a}^{(2)}$ is determined by the second order Christoffel
symbol, i.e.
\begin{equation}
D_{a}^{(2)}v^{b}=\gamma_{ac}^{(2)b}v^{c}=-\frac{1}{2}h^{bd}(D_{a}h_{cd}+D_{c}h_{ad}-D_{d}h_{ac})v^{c}.\label{eq:gamma_2nd}
\end{equation}
Then one can show
\begin{eqnarray}
\delta S & = & \frac{\sqrt{f_{c}}}{T}\int_{\mathrm{bdry}}(D_{a}t^{(2)at}+D_{a}^{(1)}t^{(1)at}+D_{a}^{(2)}t^{(0)at})\nonumber \\
 & = & \frac{\sqrt{f_{c}}}{T}\int_{\mathrm{bdry}}(\frac{1}{2}D_{d}ht^{(1)dt}+\gamma_{cd}^{(1)t}t^{(1)cd}+\frac{1}{2}h^{ab}D_{d}h_{ab}t^{(0)dt}+\gamma_{cd}^{(2)t}t^{(0)cd})\nonumber \\
 & = & \frac{\sqrt{f_{c}}}{T_{c}}\int_{\mathrm{bdry}}(\frac{1}{2}hD_{d}^{(1)}t^{(0)dt}+\gamma_{cd}^{(1)t}t^{(1)cd}+\gamma_{cd}^{(2)t}t^{(0)cd})\nonumber \\
 & = & \frac{\sqrt{f_{c}}}{T_{c}}\int_{\mathrm{bdry}}(\frac{1}{2}h[\frac{1}{2}D_{d}ht^{(0)dt}+\gamma_{ad}^{(1)t}t^{(0)ad}]+\gamma_{cd}^{(1)t}t^{(1)cd}+\gamma_{cd}^{(2)t}t^{(0)cd})\nonumber \\
 & = & \frac{\sqrt{f_{c}}}{T_{c}}\int_{\mathrm{bdry}}(\frac{1}{4}hg^{tb}[D_{d}h_{ba}+D_{a}h_{bd}-D_{b}h_{ad}]t^{(0)ad}+\gamma_{cd}^{(1)t}t^{(1)cd}+\gamma_{cd}^{(2)t}t^{(0)cd})\nonumber \\
 & = & \frac{\sqrt{f_{c}}}{T_{c}}\int_{\mathrm{bdry}}(-\frac{1}{2}h_{a}^{t}D_{d}ht^{(0)ad}-\frac{1}{4}hg^{tb}D_{b}h_{ad}t^{(0)ad}+\gamma_{cd}^{(1)t}t^{(1)cd}+\gamma_{cd}^{(2)t}t^{(0)cd})\nonumber \\
 & = & \frac{\sqrt{f_{c}}}{T_{c}}\int_{\mathrm{bdry}}(-\frac{1}{2}h_{a}^{t}D_{d}ht^{(0)ad}+\gamma_{cd}^{(1)t}t^{(1)cd}+\gamma_{cd}^{(2)t}t^{(0)cd}),\label{eq:formal}
\end{eqnarray}
where we have thrown away all the total derivative terms at most of
the steps, and employed (\ref{eq:constraint_0th}),
(\ref{eq:constraint_1st}) as well as $h_{ad}t^{(0)ad}=ph$ {(with $p$ the pressure
in equilibrium)} for our isotropic background space in the last step. Noting
\begin{eqnarray*}
t^{(1)ab}D_{a}h_{b}^{t} & \simeq & -h_{b}^{t}D_{a}t^{(1)ab}=h_{b}^{t}D_{a}^{(1)}t^{(0)ab}\\
 & = & h_{b}^{t}(\gamma_{ac}^{(1)a}t^{(0)cb}+\gamma_{ac}^{(1)b}t^{(0)ac})\\
 & = & \frac{1}{2}h_{b}^{t}D_{c}ht^{(0)cb}-\gamma_{ac}^{(2)t}t^{(0)ac}
\end{eqnarray*}
by (\ref{eq:constraint_1st}), (\ref{eq:gamma_1st}) and
(\ref{eq:gamma_2nd}), we see that (\ref{eq:formal}) is exactly the
integration of (the gravitational part of) (\ref{eq:invariant}) on
the holographic screen. {Recall that ``$\simeq$'' stands for equality
up to divergence terms on the screen, which has been used in (\ref{eq:KDh}).}

As can be seen clearly, the above discussion applies to any gravitational
theories, as long as there exists a surface term {[}like $B$ in (\ref{eq:surface}){]}
for the action functional to make the variational principle well defined.

\subsection{The case with cross-transportation}\label{cross}

When the bulk black hole background is charged ($Q\ne0$), it is known
(see e.g. \cite{coupled}) that the gravitational and electromagnetic
perturbations are coupled to each other. From the boundary point of
view, it turns out that the three kinds of transport processes are
coupled to one another, just as what happens in the thermoelectricity
phenomena. In this case, the transport coefficients form a matrix
with {off}-diagonal elements, which indicate the so-called cross-transport
processes. Now the background metric of the bulk space-time has the
general form of (\ref{eq:metric}). Let us consider the perturbation
of metric and Maxwell field\footnote{{Due to numerous indices in this subsection,
we use Latin letters to denote the bulk space-time indices within this subsection (and also Sec.\ref{sec:superfluid}).}}
\begin{eqnarray}
\tilde{g}_{ab} & = & g_{ab}+h_{ab}+q_{ab}+\cdots,\nonumber \\
F_{ab} & = & F_{ab}^{(0)}+F_{ab}^{(1)}+F_{ab}^{(2)}+\cdots,
\end{eqnarray}
where $q_{ab}$ is the second order perturbation of the metric. The perturbed Einstein tensor up to second order can be expanded as
\begin{equation}
\tilde{G}_{ab}=G_{ab}[g]+G_{ab}^{(1)}[g,h]+G_{ab}^{(1,1)}[g,h]+G_{ab}^{(0,2)}[g,q]+\cdots,
\end{equation}
where $G_{ab}[g]$ is the Einstein tensor of the background metric
$g_{ab}$; $G_{ab}^{(1)}[g,h]$ is the linearized Einstein tensor;
$G_{ab}^{(1,1)}[g,h]$ is one part of the second order perturbed Einstein
tensor which is only relevant to the first order metric perturbation
$h_{ab}$; $G_{ab}^{(0,2)}[g,q]$ is the other part of the second
order perturbed Einstein tensor which is contributed by the second
order metric perturbation $q_{ab}$. The form of $G_{ab}^{(0,2)}[g,q]$
is the same as $G_{ab}^{(1)}[g,h]$ while replacing $h_{ab}$ by $q_{ab}$.

The perturbed inverse metric up to second order is
\begin{equation}
\tilde{g}^{ab}=g^{ab}-h^{ab}+h^{ac}g_{cd}h^{db}-q^{ab}+\cdots.
\end{equation}
Explicitly, the first order perturbed Einstein tensor is
\begin{equation}
G_{ab}^{(1)}=R_{ab}^{(1)}-\frac{1}{2}R^{(0)}h_{ab}-\frac{1}{2}R_{cd}^{(0)}h^{cd}g_{ab}-\frac{1}{2}R^{(1)}g_{ab}.
\end{equation}
For the second order perturbation $R_{ab}^{(2)}$ of the Ricci tensor, we have
\begin{equation}
R_{ab}^{(2)}=R_{ab}^{(1,1)}+R_{ab}^{(0,2)},
\end{equation}
similar to the expansion of the Einstein tensor. The second order Einstein equation $G_{ab}^{(2)}+\Lambda q_{ab}=8\pi GT_{ab}^{F(2)}$
can be easily shown to have the following form:
\begin{eqnarray}
 &  & G_{ab}^{(0,2)}[g,q]+\Lambda q_{ab}\nonumber \\
 & = & [\frac{1}{2}R^{(1,1)}g_{ab}-\frac{1}{2}R_{cd}^{(1)}h^{cd}g_{ab}+\frac{1}{2}R_{cd}^{(0)}h^{ce}g_{ef}h^{fd}g_{ab}\nonumber \\
 &  & -R_{ab}^{(1,1)}+\frac{1}{2}R^{(1)}h_{ab}-\frac{1}{2}R_{cd}^{(0)}h^{cd}h_{ab}]+8\pi G T_{ab}^{F(2)}\nonumber \\
 & =: & 8\pi G(T_{ab}^{{\rm G}}+T_{ab}^{F(2)}).\label{2nd}
\end{eqnarray}

In order to investigate the holographic entropy production in this case,
we still want to construct a conserved current, but this time that turns out
to be rather subtle. Let us consider the Bianchi identity of a (fictitious) metric
$\hat{g}_{ab}=g_{ab}+q_{ab}$. It is easy to see that $\hat{\nabla}-\nabla=C_{ab}^{d}=\frac{1}{2}(\nabla_{a}q_{b}^{d}+\nabla_{b}q_{a}^{d}-\nabla^{d}q_{ab})$,
so we have
\begin{eqnarray}
0 & = & \hat{\nabla}^{a}\hat{G}_{ab}\nonumber \\
 & = & (g^{ac}-q^{ac}+\cdots)(\nabla_{c}\hat{G}_{ab}-C_{ca}^{d}\hat{G}_{db}-C_{cb}^{d}\hat{G}_{da})\nonumber \\
 & = & \nabla^{a}G_{ab}+\nabla^{a}G_{ab}^{(0,2)}[g,q]-q^{ac}\nabla_{c}G_{ab}-g^{ac}C_{ca}^{d}G_{db}-g^{ac}C_{cb}^{d}G_{da}+\cdots.
\end{eqnarray}
The second order term of the above equation is
\begin{eqnarray}
0 & = & \nabla^{a}G_{ab}^{(0,2)}[g,q]-q^{ac}\nabla_{c}G_{ab}-g^{ac}C_{ca}^{d}G_{db}-g^{ac}C_{cb}^{d}G_{da}\nonumber \\
 & = & \nabla^{a}G_{ab}^{(0,2)}[g,q]-q^{ac}\nabla_{c}G_{ab}-g^{ac}\frac{1}{2}(\nabla_{a}q_{c}^{d}+\nabla_{c}q_{a}^{d}-\nabla^{d}q_{ac})G_{db}\nonumber \\
 &  & -g^{ac}\frac{1}{2}(\nabla_{b}q_{c}^{d}+\nabla_{c}q_{b}^{d}-\nabla^{d}q_{bc})G_{da}\nonumber \\
 & = & \nabla^{a}G_{ab}^{(0,2)}[g,q]-\nabla_{c}(q^{ac}G_{ab})-\frac{1}{2}G_{cd}\nabla_{b}q^{cd}+\frac{1}{2}\nabla^{d}(q_{a}^{a}G_{db}).
\end{eqnarray}
Because the background is a stationary space-time with the time-like
Killing vector $\partial_{t}$, we consider the $t$-component of the
above equation, i.e.
\begin{eqnarray}
0 & = & (\partial_{t})^{b}[\nabla^{c}G_{bc}^{(0,2)}[g,q]-\nabla^{c}(q_{c}^{a}G_{ab})-\frac{1}{2}G_{cd}\nabla_{b}q^{cd}+\frac{1}{2}\nabla^{c}(q_{a}^{a}G_{cb})]\nonumber \\
 & = & \nabla^{c}G_{ct}^{(0,2)}[g,q]-\nabla^{c}(q_{c}^{b}G_{bt})+q_{c}^{b}G_{ab}\nabla^{c}(\partial_{t})^{a}-\frac{1}{2}G_{cd}\nabla_{t}q^{cd}+\frac{1}{2}\nabla^{c}(q_{a}^{a}G_{ct})\nonumber \\
 & = & \nabla^{a}G_{ct}^{(0,2)}[g,q]-\nabla^{c}(q_{c}^{b}G_{bt})+q_{c}^{b}G_{ab}\nabla^{c}(\partial_{t})^{a}-\frac{1}{2}\nabla_{t}(G_{cd}q^{cd})+\frac{1}{2}q_{cd}\nabla_{t}G^{cd}+\frac{1}{2}\nabla^{c}(q_{a}^{a}G_{ct})\nonumber \\
 & = & \nabla^{a}G_{ct}^{(0,2)}[g,q]-\nabla^{c}(q_{c}^{b}G_{bt})+\frac{1}{2}q^{cd}\mathcal{L}_{\partial_{t}}G_{cd}-\frac{1}{2}\nabla_{a}[G_{cd}q^{cd}(\partial_{t})^{a}]+\frac{1}{2}\nabla^{c}(q_{a}^{a}G_{ct})\nonumber \\
 & = & \nabla^{c}G_{ct}^{(0,2)}[g,q]-\nabla^{c}(q_{c}^{b}G_{bt})-\nabla_{a}[\frac{1}{2}G_{cd}q^{cd}(\partial_{t})^{a}]+\nabla^{c}(\frac{1}{2}q_{a}^{a}G_{ct}).
\end{eqnarray}
This means that $J_a:=G_{at}^{(0,2)}[g,q]-q_{a}^{b}G_{bt}-\frac{1}{2}G_{cd}q^{cd}(\partial_{t})_{a}+\frac{1}{2}q_{b}^{b}G_{at}$
is a conserved current.

{Taking into account the second order perturbed Einstein equation (\ref{2nd}),
the above conserved current can also be written as
\begin{eqnarray}
J_a & = & 8\pi G(T_{at}^{{\rm G}}+T_{at}^{F(2)})-\frac{1}{2}G_{cd}q^{cd}(\partial_{t})_{a}-q_{a}^{b}G_{bt}+\frac{1}{2}q_{b}^{b}G_{at}-\Lambda q_{at}.
\end{eqnarray}
Using the same method in the previous section, we consider the
integral of the divergence of this current and get\footnote{Here $-l$ is the outer normal
to the horizon \cite{Wald94}, which coincides with $-\partial_t$ on the horizon.}
\begin{eqnarray}
&&\int_H<-l,J>\nonumber\\
&=&\int_H (\frac{1}{2}R^{(1,1)}g_{tt}-\frac{1}{2}R_{cd}^{(1)}h^{cd}g_{tt}+\frac{1}{2}R_{cd}^{(0)}h^{ce}g_{ef}h^{fd}g_{tt}-R_{tt}^{(1,1)}+\frac{1}{2}R^{(1)}h_{tt}-\frac{1}{2}R_{cd}^{(0)}h^{cd}h_{tt}\nonumber \\
&& +8\pi G T_{tt}^{F(2)}-\frac{1}{2}G_{cd}q^{cd}g_{tt}-q_{t}^{b}G_{bt}+\frac{1}{2}q_{b}^{b}G_{tt}-\Lambda
q_{tt})\nonumber\\
&=&\int_H (-R^{(1,1)}_{ll}+8\pi G T^{F(2)}_{ll})\nonumber\\
&=&\int_{\rm bdry}<n, J>\nonumber\\
&=&\int_{\rm bdry}(\frac{1}{2}R^{(1,1)}g_{nt}-\frac{1}{2}R_{cd}^{(1)}h^{cd}g_{nt}+\frac{1}{2}R_{cd}^{(0)}h^{ce}g_{ef}h^{fd}g_{nt}-R_{nt}^{(1,1)}+\frac{1}{2}R^{(1)}h^r_{t}-\frac{1}{2}R_{cd}^{(0)}h^{cd}h^r_{t}\nonumber \\
&& +8\pi G T_{nt}^{F(2)}-\frac{1}{2}G_{cd}q^{cd}g_{nt}-q^{rb}G_{bt}+\frac{1}{2}q_{b}^{b}G_{nt}-\Lambda
q^r_{t}\nonumber\\
&=& \int_{{\rm bdry}}(-R_{nt}^{(1,1)}+8\pi G T_{nt}^{F(2)}).\label{eq:flux}
\end{eqnarray}
Here the conditions\footnote{We use ``$\heq$'' to denote an equality that holds on the horizon and ``$\
\tilde{=}\ $'' to denote an equality that holds on the boundary.\label{ft:equality}}
$h_{t\mu}\heq q_{t\mu}\heq 0$, $n_a\propto dr$ and
$g^{rt}=h^{r\mu}=q^{r\mu}=0$ are used. It is clear that the gravitational part in the
right hand side is exactly the same as in the vacuum case. Now we consider the Maxwell part.
The Maxwell part of the flux is the second order energy-momentum tensor component $T_{nt}^{F(2)}$.
Given $F^{(0)}\propto dt\wedge dr$ and that $g_{ab}$ is static,
it is easy to show that the zeroth and first orders of the energy-momentum tensor component $T_{nt}$ vanish,
so we have up to second order
\[
\int_{{\rm bdry}}T_{nt}^{F(2)}=\int_{{\rm
bdry}}{T}_{nt}=\int_{{\rm bdry}}j^{i}E_{i}.
\]
By the standard technique of the Raychaudhuri equation \cite{Wald94},
the left hand side of (\ref{eq:flux}) can be written as $T_{H}\delta S$,
so we finally obtain the general relation
\[
\delta S=\int_{\mathrm{bdry}}\Sigma\hat{\epsilon}_{[d]}.
\]
}

\section{Entropy production in holographic superconductors/superfluids}

The bulk theory of the original (and simplest) holographic superconductor/superfluid
model \cite{HHH} is a charged scalar field $\Phi$ minimally coupled
to the Maxwell field $A_{\mu}$:
\begin{equation}
I=\int[-\frac{1}{2}\nabla_{\mu}A_{\nu}F^{\mu\nu}-(\nabla_{\nu}-iA_{\nu})\Phi(\nabla^{\nu}+iA^{\nu})\Phi^{*}-m^2|\Phi|^2]\sqrt{-g}d^{d+1}x,\label{eq:HSC}
\end{equation}
in the fixed Schwarzschild-AdS black brane background
\begin{equation}
ds_{d+1}^{2}=\frac{dr^{2}}{f(r)}-f(r)dt^{2}+r^{2}d\boldsymbol{x}^{2},\qquad f(r)=\frac{r^{2}}{\ell^{2}}-\frac{2M}{r^{d-2}},\label{eq:S-AdS}
\end{equation}
i.e. without backreaction. {The case with backreaction \cite{HHH2} or even more complicated models can also be considered,
with the assumption that the equilibrium configuration is always asymptotic AdS.}

\subsection{Universal form of the holographic entropy production}

The entropy production rate (\ref{eq:production}) for various transport
processes can be covariantly written as
\begin{equation}
\Sigma(x)=-\frac{1}{T}\sum_{A}\pi_{A}{\cal L}_{\xi}\bar{\phi}^{A},\label{eq:universal}
\end{equation}
where capital Latin letters are used to index components of all
fields, and
\[
\pi_{A}(x)=\frac{1}{\sqrt{-g}}\frac{\delta I_{\mathrm{bulk}}[\bar{\phi}]}{\delta\bar{\phi}^{A}(x)}
\]
is the canonical conjugate momentum of $\bar{\phi}^{A}$, or from
the field theory point of view the expectation value of the operator
dual to ${\phi}^{A}$. Or in other words, the energy dissipation
rate is of the covariant form
\begin{equation}
E_{{\rm diss}}(x)=-\sum_{A}\pi_{A}{\cal L}_{\xi}\bar{\phi}^{A}.\label{eq:dissipation}
\end{equation}
Note that for the Maxwell field $A_{\mu}$ (or rather its boundary
components $\bar{A}_{a}$), one has
\begin{equation}
j^{a}\mathcal{L}_{\xi}\bar{A}_{a}=j^{a}(\xi^{b}D_{b}\bar{A}_{a}+\bar{A}_{b}D_{a}\xi^{b})=j^{a}\xi^{b}\bar{F}_{ba}+D_{a}(j^{a}\xi^{b}\bar{A}_{b})\label{eq:Lie_A}
\end{equation}
if the conservation $D_{a}j^{a}=0$ of the current $j^{a}$ holds,
which differs from the genuine Joule heat $j^{i}E_{i}$ by a divergence
term and thus gives the correct total energy dissipation (or entropy
production) upon integration on the whole boundary. However, if there
are charged fields in the system, then generically $D_{a}j^{a}\ne0$,
as we see explicitly later. For the moment, let us disregard this
complexity as well as the divergence term in (\ref{eq:Lie_A}).

For the holographic superconductor/superfluid model (\ref{eq:HSC}),
the entropy production corresponding to the Maxwell field $A_{\mu}$
is just included in (\ref{eq:universal}) as discussed in the previous
section, but there is also entropy production corresponding to the
scalar field $\Phi$. The latter is not a transport process in the
usual sense, and so is not a familiar entropy production process.
However, we argue that the entropy production corresponding to $\Phi$
is still given by the term with $\phi^{A}$ taken to be $\Phi$ in
(\ref{eq:universal}), i.e. the formula (\ref{eq:universal}) gives
the total entropy production rate if $\bar{\phi}^{A}$ runs over all
components of all dynamic fields in the model.%
\footnote{We would like to conjecture that the formula (\ref{eq:universal})
gives the total entropy production rate (up to divergence terms) generally,
not only for the model considered here.%
} In fact, from the thermodynamic point of view on the boundary side,
$\Pi{\cal L}_{\xi}\bar{\Phi}$ (with $\Pi$ the canonical conjugate
momentum of $\bar{\Phi}$) does be the rate of work density done on
the system. When some work is done, there is always the same amount
of energy transformed from one form to another. In general physical
systems, the energy is not necessarily transformed to heat. But in
our case (in the dual boundary system), the complete dissipation of
energy is eventually inevitable, since the system tends to settle
down and then there is no macroscopic physical degree of freedom to
``contain'' the energy. On the other hand, from the bulk point of
view, there is the same amount of energy flux going through the boundary
into the bulk and eventually being absorbed by the black brane, as
will be clear in the following discussion.%\footnote{\textcolor{blue}
%{Our argument here is only expected to hold in the linear response regime.
%The far-from-equilibrium case, such as superfluid turbulence, is highly non-trivial
%and needs further investigation.}}

In order to relate the entropy production (\ref{eq:universal}) on
the boundary to the entropy increase of the bulk black brane, one
expects that (\ref{eq:dissipation}) is the flux of some conserved
current. Actually, one may recognize (\ref{eq:dissipation}) as the
flux of the Noether current corresponding to the Killing vector field
$\xi$, i.e. the energy flux. Instead of writing down the Noether
current corresponding to the Killing vector field, however, here we
would like to give a general argument to relate $\pi_{A}{\cal L}_{\xi}\bar{\phi}^{A}$
to the flux of the current $T_{\nu}^{\mu}\xi^{\nu}$ used in the previous
section, using only the diffeomorphism invariance for a general vector
field $\xi$. Note that we will suppress the field index $A$ hereafter.

Under the diffeomorphism induced by $\xi$, the invariance of the
action (\ref{eq:HSC}) means
\begin{eqnarray}
\mathcal{L}_{\xi}I & = & \int_{{\rm bulk}}(\frac{\delta I}{\delta\phi}\mathcal{L}_{\xi}\phi+\frac{\delta I}{\delta g_{\mu\nu}}\mathcal{L}_{\xi}g_{\mu\nu})+\int_{{\rm bdry}}(\pi\mathcal{L}_{\xi}\bar{\phi}-n_{\mu}\xi^{\mu}L)\nonumber \\
 & \triangleq & \int_{{\rm bulk}}\frac{1}{2}T^{\mu\nu}\mathcal{L}_{\xi}g_{\mu\nu}+\int_{{\rm bdry}}(\pi\mathcal{L}_{\xi}\bar{\phi}-n_{\mu}\xi^{\mu}L)\nonumber \\
 & = & \int_{{\rm bulk}}T^{\mu\nu}\nabla_{\mu}\xi_{\nu}+\int_{{\rm bdry}}(\pi\mathcal{L}_{\xi}\bar{\phi}-n_{\mu}\xi^{\mu}L)\nonumber \\
 & = & \int_{{\rm bulk}}\nabla_{\mu}(T^{\mu\nu}\xi_{\nu})+\int_{{\rm bdry}}(\pi\mathcal{L}_{\xi}\bar{\phi}-n_{\mu}\xi^{\mu}L)\nonumber \\
 & = & \int_{{\rm bdry}}(n_{\mu}T^{\mu\nu}\xi_{\nu}+\pi\mathcal{L}_{\xi}\bar{\phi}-n_{\mu}\xi^{\mu}L)=0,\label{eq:integral}
\end{eqnarray}
where $L$ is the Lagrangian of (\ref{eq:HSC}), $\triangleq$ stands
for equating by the equation of motion $\frac{\delta I}{\delta\phi}=0$,
and the covariant conservation $\nabla_{\mu}T^{\mu\nu}=0$ is used
by virtue of the diffeomorphism invariance without considering the
boundary.\footnote{{There are important subtleties
for this general argument in the backreacted case, which can be remedied in a full
framework of perturbative effective action \cite{TWZ}.}} Note that we always require $\xi$ to be tangent to the
boundary, so $n_{\mu}\xi^{\mu}=0$. Taking $\xi$ as a local test
function on the boundary, one obtains
\begin{equation}
n_{\mu}T^{\mu b}\xi_{b}+\pi\mathcal{L}_{\xi}\bar{\phi}=0,\label{eq:local_flux}
\end{equation}
i.e. the local flux of the current $T_{\nu}^{\mu}\xi^{\nu}$ across
the boundary is equal to the energy dissipation rate (\ref{eq:dissipation}).
{Extending the above argument, we are allowed to investigate
more general cases of holographic entropy production and other problems in a systematic way
\cite{TWZ}.}

Then we turn to a more precise version of (\ref{eq:Lie_A}) in general
cases, using the gauge invariance
\begin{eqnarray*}
\mathcal{L}_{\Lambda}I & = & \int_{{\rm bulk}}(\frac{\delta I}{\delta\Phi}\mathcal{L}_{\Lambda}\Phi+\frac{\delta I}{\delta\Phi^{*}}\mathcal{L}_{\Lambda}\Phi^{*}+\frac{\delta I}{\delta A_{\mu}}\mathcal{L}_{\Lambda}A_{\mu})+\int_{{\rm bdry}}(\Pi\mathcal{L}_{\Lambda}\bar{\Phi}+\Pi^{*}\mathcal{L}_{\Lambda}\bar{\Phi}^{*}+j^{a}\mathcal{L}_{\Lambda}\bar{A_{a}})\\
 & \triangleq & \int_{{\rm bdry}}(i\Pi\Lambda\bar{\Phi}-i\Pi^{*}\Lambda\bar{\Phi}^{*}+j^{a}D_{a}\Lambda)=0
\end{eqnarray*}
of the action (\ref{eq:HSC}). Here $\Pi$ and $\Pi^{*}$ are the
canonical conjugate momenta of $\Phi$ and $\Phi^{*}$, respectively.
Taking $\Lambda$ as a local test function on the boundary, one obtains
\[
D_{a}j^{a}=-i\Pi\bar{\Phi}+i\Pi^{*}\bar{\Phi}^{*}
\]
locally, which means that $j^{a}$ is conserved with either Dirichlet
or Neumann boundary condition for the charged fields $\Phi$ and $\Phi^{*}$.
In the usual holographic superconductor/superfluid applications, the
Dirichlet boundary condition is taken for the charged fields, so (\ref{eq:Lie_A})
holds in this case. Otherwise, (\ref{eq:Lie_A}) becomes
\begin{eqnarray}
j^{a}\mathcal{L}_{\xi}\bar{A}_{a} & = & j^{a}\xi^{b}\bar{F}_{ba}+D_{a}(j^{a}\xi^{b}\bar{A}_{b})-\xi^{b}\bar{A}_{b}D_{a}j^{a}\nonumber \\
 & = & j^{a}\xi^{b}\bar{F}_{ba}+D_{a}(j^{a}\xi^{b}\bar{A}_{b})+\xi^{b}\bar{A}_{b}(i\Pi\bar{\Phi}-i\Pi^{*}\bar{\Phi}^{*}).\label{eq:non-conserve}
\end{eqnarray}
It is easy to see that the last term in the above equation can be
combined with $\Pi{\cal L}_{\xi}\bar{\Phi}+\Pi^{*}{\cal L}_{\xi}\bar{\Phi}^{*}$
to form a gauge invariant extension
\[
\Pi({\cal L}_{\xi}-i\bar{A}_{\xi})\bar{\Phi}+\Pi^{*}({\cal L}_{\xi}+i\bar{A}_{\xi})\bar{\Phi}^{*}
\]
of the latter, where $\bar{A}_{\xi}:=\xi^{b}\bar{A}_{b}$. The appearance
of this gauge invariant combination is expected as the scalar field
contribution in the energy dissipation rate (\ref{eq:dissipation}),
since physically the total energy dissipation should be gauge invariant,
while the Maxwell field contribution $j^{a}\xi^{b}\bar{F}_{ba}$ in
(\ref{eq:non-conserve}) is already gauge invariant. The gauge invariance
of (\ref{eq:dissipation}) is also confirmed from (\ref{eq:local_flux})
(up to divergence terms), where the energy-momentum tensor $T^{\mu\nu}$
is gauge invariant. {It is easy to see, however, that $\bar{A}_{\xi}=0$
is a rather convenient gauge choice, which means that one can simply identify the local flux
$n_{\mu}T^{F\mu b}\xi_{b}$ of the Maxwell field and that of the scalar field with $-j^{a}{\cal L}_\xi\bar{A}_{a}$
and $-(\Pi{\cal L}_{\xi}\bar{\Phi}+\Pi^{*}{\cal L}_{\xi}\bar{\Phi}^{*})$, respectively.}

\subsection{{The second order conserved current}}\label{sec:superfluid}

{In the so-called broken phase, there is a non-vanishing profile of the scalar field $\Phi$
in the (equilibrium) holographic superconductor/superfluid configurations.
In this case, the perturbations of the scalar, electromagnetic and gravitational fields are coupled to one another,
in contrast to the unbroken phase where the perturbation of the scalar field is decoupled.
However, even in the broken phase a second order conserved current can also be used to prove the entropy production formula.
In Section \ref{cross}, for the case with cross-transportation, we have already
constructed a general conserved current as}
\begin{eqnarray}
J_a=-G^{(0,2)}_{at}[g,q]+q^b_aG_{bt}+\frac{1}{2}G_{bc}q^{cb}(\partial_t)_a-\frac{1}{2}q^b_bG_{at},
\end{eqnarray}
based on the Bianchi identity. Consider the following perturbation of the Einstein-Maxwell-Scalar system:
\begin{eqnarray}
\tilde{g}_{ab}&=&g_{ab}+h_{ab}+q_{ab}+\cdots,\nonumber\\
F_{ab}&=&F^{(0)}_{ab}+F^{(1)}_{ab}+F^{(2)}_{ab}+\cdots,\nonumber\\
\Phi&=&\Phi^{(0)}+\Phi^{(1)}+\Phi^{(2)}+\cdots.
\end{eqnarray}
The second order perturbation of the Einstein equation is
\begin{eqnarray}
G^{(0,2)}+\Lambda q_{ab}=8\pi
G(T^G_{ab}+T^{F(2)}_{ab}+T^{\Phi(2)}_{ab}),
\end{eqnarray}
where $T^G_{ab}$, $T^{F(2)}_{ab}$ have been given in section III
and $T^{\Phi(2)}_{ab}$ denotes the second order perturbation of the
energy-momentum tensor of $\Phi$. We also consider the Stokes theorem
for the current $J_a$. {On the horizon, with the help of gauge, it is
\begin{eqnarray}
\int_H<-l,J>&{=}&\int_H
[-\frac{1}{8\pi G}R^{(1,1)}_{ll}+T^{F(2)}_{ll}+T^{\Phi(2)}_{ll}].%\nonumber\\
%&\heqn&\frac{1}{8\pi}\int_H\sigma^{(1)}_{ij}\sigma^{(1)ij}+Ric^{(2)}(l,l)
\end{eqnarray}
{Using the second order Raychaudhuri equation, it can be shown (see Appendix \ref{sec:Raychaudhuri}) that this
integral equals $T_H\delta S$.} On the boundary, the flux of this
current is
\begin{eqnarray}
\int_{\rm bdry}<n,J>&=&\int_{\rm bdry}[\frac{1}{8\pi G}R^{(1,1)}_{tn}-T^{F(2)}_{tn}-T^{\Phi(2)}_{tn}].
\end{eqnarray}
The integral of the Ricci part and the Maxwell part have been considered
separately in the previous section, so here we only focus on the scalar
field part. The energy-momentum tensor of $\Phi$ up to second order
is
\begin{eqnarray}
{T}^{\Phi}_{ab}&=&(\partial_a+iA_a)\Phi(\partial_b-iA_b){\Phi}^*+\mbox{c.c.}
-g_{ab}(\partial_c+iA_c)\Phi(\partial^c-iA^c){\Phi}^*\nonumber\\
&=&[\partial_a+i(A^{(0)}_a+A^{(1)}_a+A^{(2)}_a+\cdots)](\Phi^{(0)}+\Phi^{(1)}+\Phi^{(2)}+\cdots)\nonumber\\
&&\times[\partial_b-i(A^{(0)}_b+A^{(1)}_b+A^{(2)}_b+\cdots)]
({\Phi}^{(0)}+{\Phi}^{(1)}+{\Phi}^{(2)}+\cdots)^*+\mbox{c.c.}\nonumber\\
&&-\tilde{g}_{ab}[\partial_c+i(A^{(0)}_c+A^{(1)}_c+A^{(2)}_c+\cdots)](\Phi^{(0)}+\Phi^{(1)}+\Phi^{(2)}+\cdots)\nonumber\\
&&\qquad\times[\partial^c-i(A^{(0)c}+A^{(1)c}+A^{(2)c}+\cdots]({\Phi}^{(0)}+{\Phi}^{(1)}+{\Phi}^{(2)}+\cdots)^*.
\end{eqnarray}
On the boundary, the $t$-$n$ component of ${T}^{\Phi}$ is
\begin{eqnarray}
&&{T}^{\Phi}(\partial_t,n)\nonumber\\
&=&[\partial_t+i(A^{(0)}_t+A^{(1)}_t+A^{(2)}_t+\cdots)](\Phi^{(0)}+\Phi^{(1)}+\Phi^{(2)}+\cdots)\nonumber\\
&&\times[\partial_n-i(A^{(0)}_n+A^{(1)}_n+A^{(2)}_n+\cdots)]
({\Phi}^{(0)}+{\Phi}^{(1)}+{\Phi}^{(2)}+\cdots)^*+\mbox{c.c.}
\end{eqnarray}
Using the gauge choice $A_r=0$ and $A_t\ \tilde{=}\ 0$ {(as mentioned at the end of the previous subsection)},
the above equation becomes
\begin{eqnarray}
&&{T}^{\Phi}(\partial_t,n)\nonumber\\
&\tilde{=}&\partial_t(\Phi^{(0)}+\Phi^{(1)}+\Phi^{(2)}+\cdots)\partial_n
({\Phi}^{(0)}+{\Phi}^{(1)}+{\Phi}^{(2)}+\cdots)^*+\mbox{c.c.}\nonumber\\
&\tilde{=}&-(\Pi{\cal L}_\xi\bar\Phi+\Pi^*{\cal L}_\xi\bar\Phi^*).\nonumber
\end{eqnarray}
{Recall that ``$\tilde{=}$'' means equating on the boundary, as defined in Footnote \ref{ft:equality}.}
Since the background bulk space-time is stationary, we have
$\partial_t\Phi^{(0)}=0$. Then we find
\begin{eqnarray}
&&{T}^{\Phi}(\partial_t,n)\nonumber\\
&\tilde{=}&\partial_t(\Phi^{(1)}+\Phi^{(2)}+\cdots)\partial_n
({\Phi}^{(0)}+{\Phi}^{(1)}+{\Phi}^{(2)}+\cdots)^*+\mbox{c.c.}\nonumber\\
&\tilde{=}&(\partial_t\Phi^{(1)}\partial_n{\Phi}^{*(0)}
+\partial_t\Phi^{(1)}\partial_n\Phi^{*(1)}+\partial_t\Phi^{(2)}\partial_n\Phi^{*(0)}+\cdots)+\mbox{c.c.}\nonumber\\
&\tilde{=}&[\partial_t\Phi^{(1)}\partial_n\Phi^{*(1)}+\partial_t(\Phi^{(1)}\partial_n\Phi^{*(0)}+\Phi^{(2)}\partial_n{\Phi}^{*(0)})+\cdots]+\mbox{c.c.}
\end{eqnarray}
So we know up to second order,
\begin{eqnarray}
\int_{\rm bdry}{T}^{\Phi}(\partial_t,n)=\int_{\rm bdry}T^{\Phi(2)}(\partial_t,n).
\end{eqnarray}
Combining the above result with previous results, we know that the entropy
production formula also holds for the case of holographic superconductors/superfluids.

\subsection{From finite cutoff to the conformal boundary}\label{sec:conformal}

In standard AdS/CFT, as well as in AdS/CMT or AdS/QCD, the dual field
theory is defined at the conformal boundary of the asymptotic AdS
space-time. In order to clarify the entropy production in these holographic
systems, we need to extend the above discussion to the conformal boundary.
Naively, the boundary in our discussion can be put at any place, so
we may just take its position $r_{c}$ tending to the conformal infinity,
where the entropy production of the conformal field theory should
be viewed as the limit of that of the finite-cutoff theory. However,
the so-called holographic renormalization procedure (see, e.g. \cite{Skenderis})
is generically required for this limit to obtain finite physical quantities
in the dual conformal field theory. This procedure is complicated
in general, but it turns out that the total entropy production of
the dual boundary theory is not affected by this procedure, as will
be explained below.

In fact, in the holographic renormalization, one first introduces
a cutoff scale $\epsilon$ (with $\epsilon\to0$ the conformal boundary)
for the radial coordinate $z:=\frac{1}{r}$, separates the divergent
terms of physical quantities (usually the on-shell action) when $\epsilon\to0$,
and then adds a counter-term $I_{{\rm CT}}$, which is purely composed
of fields within the cutoff surface $z=\epsilon$, to the action $I$
to render it finite when on-shell. The freedom to add additional finite counter-terms
leads to different renormalization schemes. With our notations, the new action
\[
\tilde{I}=I+I_{{\rm CT}}[\bar{g}_{ab},\bar{\phi}],
\]
so (with $n_{\mu}\xi^{\mu}=0$)
\begin{eqnarray*}
\mathcal{L}_{\xi}\tilde{I} & \triangleq & \int_{{\rm bdry}}(n_{\mu}T^{\mu b}\xi_{b}+\pi\mathcal{L}_{\xi}\bar{\phi}+\frac{\delta I_{{\rm CT}}}{\delta\bar{\phi}}\mathcal{L}_{\xi}\bar{\phi}+\frac{\delta I_{{\rm CT}}}{\delta\bar{g}_{ab}}\mathcal{L}_{\xi}\bar{g}_{ab})\\
 & = & \int_{{\rm bdry}}(n_{\mu}T^{\mu b}\xi_{b}+\pi\mathcal{L}_{\xi}\bar{\phi}+\frac{\delta I_{{\rm CT}}}{\delta\bar{\phi}}\mathcal{L}_{\xi}\bar{\phi}+2\frac{\delta I_{{\rm CT}}}{\delta\bar{g}_{ab}}D_{a}\xi_{b})\\
 & = & \int_{{\rm bdry}}(n_{\mu}T^{\mu b}\xi_{b}+\tilde{\pi}\mathcal{L}_{\xi}\bar{\phi}+D_{a}\frac{2\delta I_{{\rm CT}}}{\delta\bar{g}_{ab}}\xi_{b})=0,
\end{eqnarray*}
where $\tilde{\pi}=\pi+\frac{\delta I_{{\rm CT}}}{\delta\bar{\phi}}$ is the renormalized conjugate momentum
(or more familiarly the expectation value $\langle O_\phi\rangle_{\rm CFT}$ of the operator dual to $\phi$),
and taking $\xi$ as a local test function on the boundary allows us to do the integration by parts
freely. {Recall here that $\triangleq$ stands for equating by the equation of motion, which has been used
in (\ref{eq:integral}).} Thus we obtain
\[
n_{\mu}T^{\mu b}\xi_{b}+D_{a}\frac{2\delta I_{{\rm CT}}}{\delta\bar{g}_{ab}}\xi_{b}+\tilde{\pi}\mathcal{L}_{\xi}\bar{\phi}=0.
\]
Recalling that $\xi$ should be eventually taken to be the time-like
Killing vector field on the boundary,%
\footnote{Note that if the bulk space-time is asymptotic anti-de Sitter, there
is always a boundary surface tending to the conformal infinity that
becomes the flat Minkowski space-time in the limit.%
} we know
\[
n_{\mu}T^{\mu b}\xi_{b}+D_{a}(\frac{2\delta I_{{\rm CT}}}{\delta\bar{g}_{ab}}\xi_{b})+\tilde{\pi}\mathcal{L}_{\xi}\bar{\phi}=0,
\]
i.e. the local flux of the current $T_{\nu}^{\mu}\xi^{\nu}$ across
the boundary is different from the renormalized energy dissipation
rate $-\tilde{\pi}\mathcal{L}_{\xi}\bar{\phi}$ only by a divergence
term on the cutoff surface, which gives the same total entropy production
upon integration over the whole boundary.

{In the $\epsilon\to0$ limit, another subtlety is that $\bar\phi$ (as well as the the induced boundary metric $\bar g_{ab}$ itself)
is generically vanishing or divergent. In order to obtain well-defined field quantities on the conformal boundary,
one should do a field redefinition, which can be viewed as a (constant) Weyl transformation on the boundary. Concretely,
for the holographic superconductor/superfluid model discussed here, the relevant field redefinition is
\begin{eqnarray}
% \nonumber to remove numbering (before each equation)
  g_{\mu\nu} &=& z^{-2}\hat g_{\mu\nu}, \\
  \Phi &=& z^{d-\Delta}\hat\Phi
\end{eqnarray}
with $\Delta=(d+\sqrt{d^{2}+4m^{2}\ell^{2}})/2$. Replacing the fields in the previous discussion with the above redefined fields,
the dual boundary theory now has well-defined field quantities and the entropy production is of the same form
(with the renormalized conjugate momentum $\tilde\pi$ undergoing a corresponding Weyl transformation,
including the possible conformal anomaly), while the flux of the current $T_{\nu}^{\mu}\xi^{\nu}$ has nothing to do with the redefinition.
Note that the flux of this conserved current across the horizon leads to the increase of the horizon area, as before.
The entropy production formula
\begin{equation}\label{eq:renormalized}
\Sigma=-\frac{1}{T}\sum_{A}\tilde\pi_{A}{\cal L}_{\xi}\bar{\phi}^{A}
\end{equation}
is thus justified, whose integral over the boundary space-time coincides with the entropy increase of the bulk black hole.
Remarkably, the conjugate momentum $\tilde\pi$ (the expectation value $\langle O_\phi\rangle_{\rm CFT}$) is scheme-dependent,
so is the entropy production rate (\ref{eq:renormalized}), but the total entropy production is scheme-independent.}

\section{Conclusion and discussion}

We have shown, based on a general holographic principle, the validity
of the general bulk/boundary correspondence at least at the level
of thermodynamics and hydrodynamics, where in particular, a perfect
matching between the bulk gravity and boundary system is exactly derived
for near-equilibrium entropy production on both sides by resorting
to the conserved current. Compared to the standard AdS/CFT, the bulk/boundary
correspondence discussed here is more general in the following sense.
First, the bulk space-time is not required to be asymptotically AdS {but can also be asymptotically flat or dS}.
Second, the boundary is not required to be located at the conformal
infinity {(or the asymptotic region)}. {When pushing the cutoff surface
to the conformal infinity of the asymptotic AdS space-time,
we have shown that the near-equilibrium entropy production in the simplest holographic
superconductor/superfluid model can be clearly understood. Furthermore,
we also believe that our strategy together with our statements can apply to more general spacetime with other asymptotic
behaviors such as Lifshitz or Schr\"{o}dinger, which has received much attention in AdS/CMT.}

Our boundary system, by construction, is not necessarily conformal,
so the entropy can also be produced by the bulk viscosity on the boundary
\cite{HTW}, which has been included in our discussion. It should
be noted, however, that the validity of the holographic interpretation
of the entropy production without considering bulk viscosity does
not rely on the isotropy of the background space. Or in other words,
the case with bulk viscosity requires one more constraint on the background
(equilibrium) configuration than the case without bulk viscosity.
This interesting phenomenon should be investigated further.

An important open problem is the possible holographic interpretation
of entropy production in the far-from-equilibrium case. As briefly
mentioned in the Introduction, in this case there are both
conceptual and technical difficulties for a holographic picture. For
the conceptual {side}, there is no well-established
holographic principle or dictionary in the far-from-equilibrium
case. A typical example is the debate on whether the entropy from
the bulk side corresponds to the apparent horizon or the event
horizon \cite{Rangamani,horizon}. For the technical side, the bulk
space-time dual to a far-from-equilibrium boundary system is fully
dynamic, which is difficult to explore analytically. {However, some interesting
analytic works have been done along a similar direction \cite{G J,GJ2}.}
{We hope to come back to this problem later.}

\begin{acknowledgments}
{We thank Sijie Gao, Yi Ling, and Junbao Wu for helpful
discussions. itriou, and Tassos Petkou for their valuable
discussions.This work is partly supported by the National
Natural Science Foundation of China (Grant Nos. 11175245 and 11475179).
It is also supported in part by the Belgian Federal
Science Policy Office through the Interuniversity Attraction Pole
P7/37, by FWO-Vlaanderen through the project
G020714N, and by the Vrije Universiteit Brussel through the
Strategic Research Program ``High-Energy Physics". H.Z. is also an individual FWO fellow supported by 12G3515N. He would also like to thank the organizers of the  long term program "Quantum Gravity, Black Holes and Strings" at KITPC for the fantastic infrastructure they provide and the generous financial support they of they offer, which speeds this project.}
\end{acknowledgments}
\appendix

\section{Black-hole thermodynamics with the topological charge\label{sec:thermodynamics}}

In this appendix, we will consider the black-hole thermodynamics with
the topological charge $\varepsilon$ in the general Lovelock-Maxwell
theory, following the formulation proposed in Ref.\cite{T W}. In
this formulation, we assume a standard form
\[
ds_{d+1}^{2}=\frac{dr^{2}}{f(r)}-f(r)dt^{2}+r^{2}d\Omega_{d-1}^{(k)2}
\]
of the metric and focus on an ``equal-potential'' surface $f={\rm const}$,
and reinterpret the Lovelock equations of motion as a generalized
first law, which gives the traditional black-hole first law in the
case $f=0$ (the horizon).

As described in Ref.\cite{T W}, we can read from the generalized
first law the ADM mass
\begin{equation}
M=\frac{d-1}{16\pi}\Omega_{d-1}\left(r^{d}\sum_{j}\tilde{\alpha}_{j}(\frac{k-f}{r^{2}})^{j}+\frac{Q^{2}}{r^{d-2}}\right)\label{eq:M}
\end{equation}
and generalized Wald entropy
\[
S=\frac{d-1}{4}\Omega_{d-1}r^{d-1}\sum_{j}\frac{\tilde{\alpha}_{j}j}{d-2j+1}(\frac{k-f}{r^{2}})^{j-1}
\]
of the maximally symmetric charged black hole, where $\tilde{\alpha}_{j}$
is proportional to the $j$-th Lovelock coupling constant. Viewing
$M$ as a function of $(f,Q,k)$ or $(r,Q,k)$, differentiation of
(\ref{eq:M}) gives
\begin{equation}
dM=-\frac{f'}{4\pi}\frac{\partial M}{\partial f}4\pi dr+\frac{\partial M}{\partial Q}dQ+\frac{d-1}{16\pi}\Omega_{d-1}r^{d-2}\sum_{j}\tilde{\alpha}_{j}j(\frac{k-f}{r^{2}})^{j-1}dk.\label{eq:dM}
\end{equation}
However, we need the differentiation of $M$ as a function of $(S,Q,k)$
to obtain the generalized first law with the topological charge, so
we should consider the substitution
\[
(r,k)\to(S,k)
\]
of variables. In fact, we have%
\footnote{Here we need the relation $\frac{\partial S}{\partial r}=-4\pi\frac{\partial M}{\partial f}$
discovered in Ref.\cite{T W}.%
}
\[
\left(\begin{array}{cc}
\frac{\partial r}{\partial S} & \frac{\partial k}{\partial S}\\
\frac{\partial r}{\partial k} & \frac{\partial k}{\partial k}
\end{array}\right)=\left(\begin{array}{cc}
\frac{\partial S}{\partial r} & \frac{\partial k}{\partial r}\\
\frac{\partial S}{\partial k} & \frac{\partial k}{\partial k}
\end{array}\right)^{-1}=\left(\begin{array}{cc}
-4\pi\frac{\partial M}{\partial f} & 0\\
\frac{\partial S}{\partial k} & 1
\end{array}\right)^{-1}=\left(\begin{array}{cc}
-(4\pi\frac{\partial M}{\partial f})^{-1} & 0\\
(4\pi\frac{\partial M}{\partial f})^{-1}\frac{\partial S}{\partial k} & 1
\end{array}\right),
\]
which leads to
\[
dr=\frac{\partial r}{\partial S}dS+\frac{\partial r}{\partial k}dk=-(4\pi\frac{\partial M}{\partial f})^{-1}dS+(4\pi\frac{\partial M}{\partial f})^{-1}\frac{\partial S}{\partial k}dk.
\]
Substitution of the above equation into (\ref{eq:dM}) gives
\begin{eqnarray*}
dM & = & \frac{f'}{4\pi}dS+\varphi dQ+[\frac{d-1}{16\pi}\Omega_{d-1}r^{d-2}\sum_{j}\tilde{\alpha}_{j}j(\frac{k-f}{r^{2}})^{j-1}+(4\pi\frac{\partial M}{\partial f})^{-1}\frac{\partial S}{\partial k}(-f'\frac{\partial M}{\partial f})]dk\\
 & = & \frac{f'}{4\pi}dS+\varphi dQ+\Omega_{d-1}[\frac{d-1}{16\pi}r^{d-2}\sum_{j}\tilde{\alpha}_{j}j(\frac{k-f}{r^{2}})^{j-1}-\frac{f'}{4\pi}\frac{d-1}{4}r^{d-3}\sum_{j}\frac{\tilde{\alpha}_{j}j(j-1)}{d-2j+1}(\frac{k-f}{r^{2}})^{j-2}]dk\\
 & = & \frac{f'}{4\pi}dS+\varphi dQ+\frac{\Omega_{d-1}k^{\frac{3-d}{2}}}{8\pi}[r^{d-2}\sum_{j}\tilde{\alpha}_{j}j(\frac{k-f}{r^{2}})^{j-1}-f'r^{d-3}\sum_{j}\frac{\tilde{\alpha}_{j}j(j-1)}{d-2j+1}(\frac{k-f}{r^{2}})^{j-2}]dk^{\frac{d-1}{2}}\\
 & = & T_{UV}dS+\varphi dQ+\frac{\varepsilon^{\frac{3-d}{d-1}}S'}{2\pi(d-1)}d\varepsilon
\end{eqnarray*}
with $T_{UV}=\frac{f'}{4\pi}$ the so-called Unruh-Verlinde temperature%
\footnote{This temperature tends to the Hawking temperature of the black hole
when the holographic screen approaches the horizon.%
}, $\varphi=\frac{\partial M}{\partial Q}$ the electric potential
and $\varepsilon=k^{\frac{d-1}{2}}$ the topological charge. In comparison
to (\ref{eq:1st_law_plus}), the chemical potential $\mu$ in (\ref{eq:chemical})
is just the difference of $\varphi$ between the horizon and the holographic
screen (up to a redshift factor). Thus it is natural to conjecture
that the conjugate quantity $\varsigma$ {[}of the form (\ref{eq:sigma})
in the Einstein-Maxwell theory{]} is just the difference of
\[
\varpi=\frac{\varepsilon^{\frac{3-d}{d-1}}S'}{2\pi(d-1)}
\]
between the horizon and the holographic screen (up to a redshift factor),
which is indeed the case in the Einstein-Maxwell theory. But it is
still unclear whether $\varpi$ can be viewed as some kind of potential.

\section{The increase of horizon area from the Raychaudhuri equation\label{sec:Raychaudhuri}}

In this appendix, we will show that the entropy increase of the bulk black hole is equal
to the total entropy production on the boundary system, i.e.
\begin{eqnarray}
T_H\frac{\delta A_H}{4G}=\int_{\rm bdry}\Sigma,
\end{eqnarray}
where $\Sigma$ is defined in Eq.(\ref{eq:rate}). In Section III, we
have shown that the flux of the conserved current $-\frac{1}{8\pi
G}\ G_{\mu t}^{(1,1)}$ on the holographic screen is just the entropy
production of the boundary system. In this section, we will show that the
total flux of the same current on the horizon is equal to $T_H\frac{\delta
A_H}{4G}$.

From Eq.(\ref{eq:correspond}), the integral on the horizon is
\begin{eqnarray}
\int_HT^{\mu}_t\lambda_{\mu}\tilde{\epsilon}_{[d]}=-\frac{1}{8\pi G}\int_HR^{(1,1)}_{ll}\tilde{\epsilon}_{[d]}.
\end{eqnarray}
The concrete form of $R^{(1,1)}_{ll}$ is
\begin{eqnarray}
R^{(1,1)}_{ll}
&\hat{=}&-\partial_{u}\left[\frac{1}{2}h^{\rho\sigma}(h_{\rho\sigma,u}+h_{u\sigma,\rho}-h_{u\rho,\sigma})\right]
+\partial_{u}\left[\frac{1}{2}h^{u\sigma}(h_{\alpha\sigma,u}+h_{u\sigma,\alpha}-h_{u\alpha,\sigma})\right]\nonumber\\
&&+\partial_{i}\left[\frac{1}{2}h^{i\sigma}(h_{u\sigma,u}+h_{u\sigma,u}-h_{uu,\sigma})\right]
+\partial_{r}\left[\frac{1}{2}h^{r\sigma}(h_{u\sigma,u}+h_{u\sigma,u}-h_{uu,\sigma})\right]\nonumber\\
&&-\frac{1}{2}g^{\rho\sigma}(h_{u\sigma,\eta}+h_{\eta\sigma,u}-h_{u\eta,\sigma})
\frac{1}{2}g^{\eta\lambda}(h_{\rho\lambda,u}+h_{u\lambda,\rho}-h_{u\rho,\lambda})\nonumber\\
&&-\frac{1}{2}g^{\rho\sigma}(h_{u\sigma,\eta}+h_{\eta\sigma,u}-h_{u\eta,\sigma})
\frac{1}{2}h^{\eta\lambda}(g_{\rho\lambda,u}+g_{u\lambda,\rho}-g_{u\rho,\lambda})\nonumber\\
&&-\frac{1}{2}h^{\rho\sigma}(g_{u\sigma,\eta}+g_{\eta\sigma,u}-g_{u\eta,\sigma})
\frac{1}{2}g^{\eta\lambda}(h_{\rho\lambda,u}+h_{u\lambda,\rho}-h_{u\rho,\lambda})\nonumber\\
&&-\frac{1}{2}h^{\rho\sigma}(g_{u\sigma,\eta}+g_{\eta\sigma,u}-g_{u\eta,\sigma})
\frac{1}{2}h^{\eta\lambda}(g_{\rho\lambda,u}+g_{u\lambda,\rho}-g_{u\rho,\lambda})\nonumber\\
&&-\frac{1}{2}h^{\rho\sigma}(h_{u\sigma,\eta}+h_{\eta\sigma,u}-h_{u\eta,\sigma})
\frac{1}{2}g^{\eta\lambda}(g_{\rho\lambda,u}+g_{u\lambda,\rho}-g_{u\rho,\lambda})\nonumber\\
&&-\frac{1}{2}g^{\rho\sigma}(g_{u\sigma,\eta}+g_{\eta\sigma,u}-g_{u\eta,\sigma})
\frac{1}{2}h^{\eta\lambda}(h_{\rho\lambda,u}+h_{u\lambda,\rho}-h_{u\rho,\lambda})\nonumber\\
&&+\frac{1}{2}g^{\rho\sigma}(h_{\rho\sigma,\eta}+h_{\eta\sigma,\rho}-h_{\eta\rho,\sigma})
\frac{1}{2}g^{\eta\lambda}(h_{u\lambda,u}+h_{u\lambda,u}-h_{uu,\lambda})\nonumber\\
&&+\frac{1}{2}g^{\rho\sigma}(h_{\rho\sigma,\eta}+h_{\eta\sigma,\rho}-h_{\eta\rho,\sigma})
\frac{1}{2}h^{\eta\lambda}(g_{u\lambda,u}+g_{u\lambda,u}-g_{uu,\lambda})\nonumber\\
&&+\frac{1}{2}h^{\rho\sigma}(g_{\rho\sigma,\eta}+g_{\eta\sigma,\rho}-g_{\eta\rho,\sigma})
\frac{1}{2}g^{\eta\lambda}(h_{u\lambda,u}+h_{u\lambda,u}-h_{uu,\lambda})\nonumber\\
&&+\frac{1}{2}h^{\rho\sigma}(g_{\rho\sigma,\eta}+g_{\eta\sigma,\rho}-g_{\eta\rho,\sigma})
\frac{1}{2}h^{\eta\lambda}(g_{u\lambda,u}+g_{u\lambda,u}-g_{uu,\lambda})\nonumber\\
&&+\frac{1}{2}h^{\rho\sigma}(h_{\rho\sigma,\eta}+h_{\eta\sigma,\rho}-h_{\eta\rho,\sigma})
\frac{1}{2}g^{\eta\lambda}(g_{u\lambda,u}+g_{u\lambda,u}-g_{uu,\lambda})\nonumber\\
&&+\frac{1}{2}g^{\rho\sigma}(g_{\rho\sigma,\eta}+g_{\eta\sigma,\rho}-g_{\eta\rho,\sigma})
\frac{1}{2}h^{\eta\lambda}(h_{u\lambda,u}+h_{u\lambda,u}-h_{uu,\lambda})\label{R11}.
\end{eqnarray}
Using the gauge in section III, $h^{r\mu}\ =\ 0$, it is easy to see
$h_{u\mu}\heq 0$. Now we need to calculate all terms in above
equation. {Recall that ``$\hat{=}$'' means equating on the horizon, as defined in Footnote \ref{ft:equality}.}

Obviously, the first and second terms are zero because we can do
time integral to make them to be boundary term. The boundary term
vanish because of the zero initial data and the Price
law \cite{Pricelaw}.

The third term can be expressed as
\begin{eqnarray}
&&\partial_{i}\left[\frac{1}{2}h^{i\sigma}(h_{u\sigma,u}+h_{u\sigma,u}-h_{uu,\sigma})\right]\nonumber\\
&\hat{=}&\partial_i\left[\frac{1}{2}h^{ir}(-h_{uu,r})\right]\nonumber\\
&\hat{=}&0.
\end{eqnarray}
In the first step, we have used the fact $h_{u\mu}\heq 0$. In the second
step, we have used the gauge condition $h^{ir}=0$.

The fourth term is zero because the gauge condition $h^{r\mu}=0$.

Let us consider the non-derivative terms. The first non-derivative
term in Eq.(\ref{R11}) is
\begin{eqnarray}
&&\frac{1}{2}g^{\rho\sigma}(h_{u\sigma,\eta}+h_{\eta\sigma,u}-h_{u\eta,\sigma})
\frac{1}{2}g^{\eta\lambda}(h_{\rho\lambda,u}+h_{u\lambda,\rho}-h_{u\rho,\lambda})\nonumber\\
&\hat{=}&\frac{1}{2}g^{\rho\sigma}h_{u\sigma,\eta}\frac{1}{2}g^{\eta\lambda}h_{\rho\lambda,u}
+\frac{1}{2}g^{\rho\sigma}h_{\eta\sigma,u}\frac{1}{2}g^{\eta\lambda}h_{\rho\lambda,u}
-\frac{1}{2}g^{\rho\sigma}h_{u\eta,\sigma}\frac{1}{2}g^{\eta\lambda}h_{\rho\lambda,u}\nonumber\\
&&+\frac{1}{2}g^{\rho\sigma}h_{u\sigma,\eta}\frac{1}{2}g^{\eta\lambda}h_{u\lambda,\rho}
+\frac{1}{2}g^{\rho\sigma}h_{\eta\sigma,u}\frac{1}{2}g^{\eta\lambda}h_{u\lambda,\rho}
-\frac{1}{2}g^{\rho\sigma}h_{u\eta,\sigma}\frac{1}{2}g^{\eta\lambda}h_{u\lambda,\rho}\nonumber\\
&&-\frac{1}{2}g^{\rho\sigma}h_{u\sigma,\eta}\frac{1}{2}g^{\eta\lambda}h_{u\rho,\lambda}
-\frac{1}{2}g^{\rho\sigma}h_{\eta\sigma,u}\frac{1}{2}g^{\eta\lambda}h_{u\rho,\lambda}
+\frac{1}{2}g^{\rho\sigma}h_{u\eta,\sigma}\frac{1}{2}g^{\eta\lambda}h_{u\rho,\lambda}\nonumber\\
&\hat{=}&\frac{1}{2}g^{\rho\sigma}h_{u\sigma,r}\frac{1}{2}g^{r\lambda}h_{\rho\lambda,u}
+\frac{1}{2}g^{\rho\sigma}h_{\eta\sigma,u}\frac{1}{2}g^{\eta\lambda}h_{\rho\lambda,u}
-\frac{1}{2}g^{\rho r}h_{u\eta,r}\frac{1}{2}g^{\eta\lambda}h_{\rho\lambda,u}\nonumber\\
&&+\frac{1}{2}g^{\rho\sigma}h_{u\sigma,r}\frac{1}{2}g^{r\lambda}h_{u\lambda,\rho}
+\frac{1}{2}g^{r\sigma}h_{\eta\sigma,u}\frac{1}{2}g^{\eta\lambda}h_{u\lambda,r}
-\frac{1}{2}g^{r\sigma}h_{u\eta,\sigma}\frac{1}{2}g^{\eta\lambda}h_{u\lambda,r}\nonumber\\
&&-\frac{1}{2}g^{\rho\sigma}h_{u\sigma,r}\frac{1}{2}g^{r\lambda}h_{u\rho,\lambda}
-\frac{1}{2}g^{\rho\sigma}h_{\eta\sigma,u}\frac{1}{2}g^{\eta
r}h_{u\rho,r}
+\frac{1}{2}g^{\rho\sigma}h_{u\eta,\sigma}\frac{1}{2}g^{\eta r}h_{u\rho,r}\nonumber\\
&\hat{=}&\frac{1}{2}g^{\rho\sigma}h_{\eta\sigma,u}\frac{1}{2}g^{\eta\lambda}h_{\rho\lambda,u}
+\frac{1}{2}g^{ru}h_{uu,r}\frac{1}{2}g^{ru}h_{uu,r}
+\frac{1}{2}g^{ur}h_{uu,r}\frac{1}{2}g^{ur}h_{uu,r}\nonumber\\
&\hat{=}&\frac{1}{2}g^{\rho\sigma}h_{\eta\sigma,u}\frac{1}{2}g^{\eta\lambda}h_{\rho\lambda,u}
+\frac{1}{2}(h_{uu,r})^2\nonumber\\
&\hat{=}&\frac{1}{2}g^{ji}h_{qi,u}\frac{1}{2}g^{qk}h_{jk,u}
+\frac{1}{2}(h_{uu,r})^2\nonumber\\
&\hat{=}&\frac{1}{d-1}(\theta^{(1)})^2+(\sigma^{(1)})^2+\frac{1}{2}(h_{uu,r})^2,
\end{eqnarray}
where Greek indices run from $0$ to $d$ and Latin indices run from
$1$ to $d-1$. In the second line, we have used $h_{u\mu}\heq 0$. In the
following three steps, we have used the gauge $h^{r\mu}=0$, $h_{u\mu}\heq
0$, $g_{ur}\heq g^{ur}\heq 1$, $g_{uu}\heq g_{ui}\heq g_{rr}\heq
g_{ri}\heq 0$ and $g^{uu}\heq g^{ui}\heq g^{rr}\heq g^{ri}\heq 0$.
In the last step, we have used the definition
$\frac{1}{2}h_{ij,u}=:\frac{1}{d-1}\theta^{(1)}g_{ij}+\sigma^{(1)}_{ij}$.

Using similar analysis, one can get
\begin{eqnarray}
R^{(1,1)}_{ll}&\hat{=}&
-\frac{1}{d-1}(\theta^{(1)})^2-(\sigma^{(1)})^2+\frac{1}{2}(h_{uu,r})^2-\theta^{(1)}h_{uu,r}-\partial_u\left[\frac{1}{4}h^{ij}h_{ij}g_{uu,r}\right]\nonumber\\
&&-\partial_{u}\left[\frac{1}{2}h^{\rho\sigma}(h_{\rho\sigma,u}+h_{u\sigma,\rho}-h_{u\rho,\sigma})\right]\nonumber\\
&&+\partial_{u}\left[\frac{1}{2}h^{u\sigma}(h_{\alpha\sigma,u}+h_{u\sigma,\alpha}-h_{u\alpha,\sigma})\right].\label{Rll2}
\end{eqnarray}
In order to consider the value of $R^{(1,1)}_{ll}$ on the horizon, we
need the value of first order perturbation of horizon expansion
$\theta^{(1)}$. To do this, we need to consider the perturbation of
Raychaudhuri equation. The first order perturbation of this equation
is
\begin{eqnarray}
\dot{\theta}^{(1)}&=&\kappa^{(0)}\theta^{(1)}-\theta^{(0)}\theta^{(1)}-2\sigma^{(0)}\cdot\sigma^{(1)}-R^{(1,1)}_{ll}\nonumber\\
&=&\kappa^{(0)}\theta^{(1)},\label{T4}
\end{eqnarray}
where we have used the fact $\theta^{(0)}\heq\sigma^{(0)}\heq 0$ and the linearized
vacuum Einstein equation $R^{(1,1)}_{ll}=0$. For the non-vacuum case, one
needs to consider the concrete form of $R^{(1,1)}_{ll}$. For the
Einstein-Maxwell-Scalar system, the linearized Einstein equation
implies
\begin{eqnarray}
R^{(1,1)}_{ll}=8\pi G F_{li}^{(0)}F_l^{(1)i}+8\pi G\del_l\Phi^{(0)}\del_l\Phi^{(1)}.
\end{eqnarray}
Because the back ground is a stationary black hole, one can show
that $F_{li}^{(0)}\heq\del_l\Phi^{(0)}\heq 0$ based on the zeroth order
Raychaudhuri equation. This means that $R^{(1,1)}_{ll}\heq 0$ also holds for the
Einstein-Maxwell-Scalar system, so as Eq.(\ref{T4}). Eq.(\ref{T4})
is an ordinary differential equation on the horizon. With the zero
initial data, one can get $\theta^{(1)}\heq 0$, so Eq.(\ref{Rll2})
becomes
\begin{eqnarray}
R^{(1,1)}_{ll}&\hat{=}&-(\sigma^{(1)})^2-\frac{1}{2}(h_{uu,r})^2-\partial_u\left[\frac{1}{4}h^{ij}h_{ij}g_{uu,r}\right]\nonumber\\
&&-\partial_{u}\left[\frac{1}{2}h^{\rho\sigma}(h_{\rho\sigma,u}+h_{u\sigma,\rho}-h_{u\rho,\sigma})\right]\nonumber\\
&&+\partial_{u}\left[\frac{1}{2}h^{u\sigma}(h_{\alpha\sigma,u}+h_{u\sigma,\alpha}-h_{u\alpha,\sigma})\right].\label{Rll3}
\end{eqnarray}
Taking a suitable coordinate transformation, one can show that
$(h_{uu,r})^2$ vanishes on the horizon. We have shown that the first
order perturbation $\theta^{(1)}$ of the horizon expansion vanishes. This means that the
variation of horizon area is contributed by the second order
perturbation. The second order Raychaudhuri equation is
\begin{eqnarray}
\dot{\theta}^{(2)}-\kappa^{(0)}\theta^{(2)}&=&-\sigma^{(1)}_{ij}\sigma^{(1)ij}-R^{(2)}_{ll}\nonumber\\
&=&-\sigma^{(1)}_{ij}\sigma^{(1)ij}-8\pi G T^{(2)}_{ll},
\end{eqnarray}
where the second order Einstein equation is used. Based on Wald's standard technique \cite{G W,Wald94}, we know that
\begin{eqnarray}
\int_H\dot{\theta}^{(2)}-\kappa^{(0)}\theta^{(2)}=-\kappa^{(0)}\delta
A_H=-\int_H[(\sigma^{(1)})^2+8\pi G T^{(2)}_{ll}].
\end{eqnarray}
Combining the above equation with Eq.(\ref{Rll3}), it is easy to see that
up to second order perturbation
\begin{eqnarray}
T_H\delta S&=&\frac{1}{8\pi G}\int_H[-R^{(1,1)}_{ll}+8\pi G
T^{(2)}_{ll}]\nonumber\\
&=&\frac{1}{8\pi G}\int_{\rm bdry}[-R^{(1,1)}_{nt}+8\pi G
T^{(2)}_{nt}]\nonumber\\
&=&\int_{\rm bdry}\Sigma.
\end{eqnarray}

\end{document}